\documentclass[]{pasj01}

\usepackage{bm}
\newcommand{\mdotinave}{180}
\newcommand{\mdotoutave}{0.13}
\newcommand{\mdotoutavedisk}{4.6\times10^{-6}}
\newcommand{\mdotoutend}{24}
\newcommand{\rqss}{600}
\newcommand{\rtrap}{270}
\newcommand{\rtrapf}{450}
\newcommand{\rtrapsph}{330}
\newcommand{\rinflow}{40}
\newcommand{\rlau}{280}
\newcommand{\rlauinf}{190}
\newcommand{\rlaucyl}{230}
\newcommand{\rlauinfcyl}{140}

\Received{2020/10/31}
\Accepted{2021/01/20}
 
\begin{document} 

\title{ 
  Outflow from super-Eddington flow:
  where it originates from and how much impact it gives?
}

\author{Takaaki \textsc{KITAKI}\altaffilmark{1}}
\altaffiltext{1}{Department of Astronomy, Graduate School of Science, Kyoto University, Kitashirakawa-Oiwake-cho, Sakyo-ku, Kyoto 606-8502, Japan}
\email{kitaki@kusastro.kyoto-u.ac.jp}

\author{Shin \textsc{Mineshige}\altaffilmark{1}}

\author{Ken \textsc{Ohsuga}\altaffilmark{2}}
\altaffiltext{2}{Center for Computational Sciences, University of Tsukuba, Ten-nodai, 1-1-1 Tsukuba, Ibaraki 305-8577, Japan}

\author{Tomohisa \textsc{Kawashima}\altaffilmark{3}}
\altaffiltext{3}{Institute for Cosmic Ray Research, The University of Tokyo, 5-1-5 Kashiwanoha, Kashiwa, Chiba 277-8582, Japan}


\KeyWords{accretion, accretion disks --- radiation: dynamics --- stars: black holes}

\maketitle

\begin{abstract}
  It is widely believed that super-Eddington accretion flow can produce powerful outflow, but 
  where it originates from and how much mass and energy are carried away to which directions?
  To answer to these questions, we newly perform a large-box, two-dimensional radiation hydrodynamic simulation,
  paying special attention lest the results should depend on adopted initial and boundary conditions. 
  We could achieve a quasi-steady state in an unprecedentedly large range, $r=2~r_{\rm S}$--$600~r_{\rm S}$
  (with $r_{\rm S}$ being the Schwarzschild radius) from the black hole.
  The accretion rate onto the central $10 ~M_{\odot}$ black hole is $\dot{M}_{\rm BH} \sim \mdotinave ~L_{\rm Edd}/c^{2}$,
  whereas the mass outflow rate is ${\dot M}_{\rm outflow} \sim \mdotoutend ~L_{\rm Edd}/c^2$
  (where $L_{\rm Edd}$ and $c$ are the Eddington luminosity and the speed of light, respectively).
  The ratio (${\dot M}_{\rm outflow}/{\dot M}_{\rm BH} \sim 0.14$) is much less than those reported previously.
  By careful inspection we find that most of outflowing gas which reach the outer boundary
  originates from the region at $R\lesssim\rlauinfcyl~r_{\rm S}$, 
  while gas at $\rlauinfcyl~r_{\rm S}$--$230 ~r_{\rm S}$ forms failed outflow. 
  Therefore, significant outflow occurs inside the trapping radius $\sim \rtrapf ~r_{\rm S}$.
  The mechanical energy flux (or mass flux) reaches its maximum in the direction of 
  $\sim 15^\circ$ ($\sim 80^\circ$) from the rotation axis.
  The total mechanical luminosity is $L_{\rm mec}\sim  0.16~L_{\rm Edd}$,
  while the isotropic X-ray luminosity varies from $L_{\rm X}^{\rm ISO}\sim 2.9~L_{\rm Edd}$, (for a face-on observer) to $\sim 2.1~L_{\rm Edd}$ (for a nearly edge-on observer).
  The power ratio is $L_{\rm mec}/L_{\rm X}^{\rm ISO}\sim 0.05$--$0.08$, in good agreement with the observations of Ultra-Luminous X-ray sources surrounded by optical nebulae.
\end{abstract}

\section{Introduction}
\label{sec-introduction}
It is well known that gas accretion onto a black hole produce enormous energy, thereby giving rise to a variety of active phenomena in black hole objects, such as X-ray binaries and Active Galactic Nuclei (AGNs).
The gas in accretion disk falls onto the central black hole via transportation of the angular momentum by the viscosity,
and releases the gravitational energy in forms of the radiation energy
and/or the mechanical energy (see e.g., Shakura \& Sunyaev 1973).
There is a classical limit to the total amount of the radiation energy released per unit time by accretion; that is what we call the Eddington luminosity, $L_{\rm Edd}$.
It is derived from the balance between the radiation force and the gravitational force under spherical symmetry, and is written as,
\begin{eqnarray}
  L_{\rm Edd}&\equiv&\frac{4\pi cGM_{\rm BH}}{\kappa_{\rm es}}\simeq 1.26\times10^{39}\left(\frac{M_{\rm BH}}{10M_{\odot}}\right)~{\rm erg}~{\rm s^{-1}}.
\end{eqnarray}
Here, $c$ is the light speed, $G$ is the gravitational constant, $M_{\rm BH}$ is the mass of a central black hole, $\kappa_{\rm es}$ is the Thomson scattering opacity,
and we assumed the hydrogen abundance of $X = 1.0$.

It is now widely accepted that the classical limit can be exceeded in disk accretion because of the separation of the directions of the gas inflow and of the radiation output.
The super-Eddington accretion flow is the gas flow with an extremely high accretion rates, $\dot{M}_{\rm BH}\gg L_{\rm Edd}/c^{2}$,
and is known to shine at the super-Eddington luminosity.
There are several astrophysical objects that are known to harbor super-Eddington accretors:
Good candidates are Ultra-Luminous X-ray sources (ULXs), and some of microquasars (e.g., GRS1915+105) and narrow-line Seyfert 1 galaxies (NLS1s: e.g. Mineshige et al. 2000; Jin et al. 2017).

The ULXs are bright X-ray compact sources whose X-ray luminosity is $10^{39}$--$10^{41}$ [erg/s] and have been discovered in off-nuclear regions of nearby galaxies (Kaaret et al. 2017 for a recent review).
There are two main ideas to explain high luminosity:
one is the sub-Eddington accretion onto the intermediate-mass black holes (IMBH: Makishima et al. 2000; Miller et al. 2004),
and another is the super-Eddington accretion onto the stellar mass black holes (Watarai et al. 2001; King et al. 2001).

The situation had drastically changed after the discovery of the so-called ULX pulsars as a subgroup of ULXs but showing periodic X-ray pulses
(e.g., M82 X-2, Bachetti et al. 2014;
NGC7793 P13, F\"{u}rst et al. 2016, Israel et al. 2017).
Now they are known to possess magnetized neutron stars.
These discoveries support the super-Eddington accretion scenario,
although there still remains a room for the IMBH hypothesis to survive to account for extremely high luminosity of the HLX (see, e.g. Barrows et al. 2019 and references therein).

In the theoretical aspects, 
one of the most prominent features of the super-Eddington accretion flow is the photon-trapping effect (Katz 1977; Begelman 1978; Abramowicz et al. 1988).
When the mass accretion rate is very large, $\dot{M}\gg L_{\rm Edd}/c^{2}$,
so is the vertical optical depth, $\tau_{\rm e}$ ($=\kappa_{\rm es}\Sigma$ with $\Sigma$ being the surface density),
since we have $\dot{M} =2 \pi r \Sigma |v_{r}|$ (with $v_{r}$ being the radial velocity).
Then, the photon diffusion timescale ($\propto\tau_{\rm e}$) to the disk surface can exceed the accretion timescale ($=r/|v_{r}|$).
When this occurs,
photons will be trapped within gas flow and swallowed by a central black hole together with gas.
The photon-trapping radius $R_{\rm trap}$ inside which the photon-trapping effect is significant is given by (e.g. Kato et al. 2008),
\begin{eqnarray}
  R_{\rm trap}&=&\frac{3}{2}\frac{H}{R}\dot{m}_{\rm BH}r_{\rm S}. \label{eq-trap}
\end{eqnarray}
Here, $H$ is the scale-height of the accretion disk,
$R$ is the radius in the cylindrical coordinates,
$\dot{m}_{\rm BH}\equiv\dot{M}_{\rm BH}/(L_{\rm Edd}/c^{2})$ is the normalized mass accretion rate onto the black hole, 
and $r_{\rm s}\equiv 2GM_{\rm BH}/c^{2}$ is the Schwarzschild radius.

Outflow is another prominent feature of the super-Eddington accretion flow.
The super-Eddington luminosity implies that the radiation force is greater than the gravitational force,
leading to the emergence of radiation-pressure driven outflow (Shakura \& Sunyaev 1973).
Once gas is blown away from the disk surface in a form of outflow,
it will inevitably give impact on the environments far from the central black hole.
The accretion disk structure itself should also be affected,
since the accretion rate within the disk is no longer constant in space. 
Moreover, the emergent spectrum will be modified via the Comptonization by the outflowing gas (Kawashima et al. 2012; Kitaki et al. 2017; Narayan et al. 2017). 
This will account for the observed spectra in the so-called ultra-luminous state of ULXs (Gladstone et al. 2009; Kawashima et al. 2012).

We can derive the launching radius by the considerations of the balance between the radiation force and the gravitational force at the disk surface; that is
\begin{eqnarray}
  R_{\rm lau}&\sim& A_{\rm lau}\dot{m}_{\rm BH} r_{\rm S}. \label{eq-launching}
\end{eqnarray}
Here, $A_{\rm lau}$ is a constant of order unity, depending on the geometry.
The launching radius is essentially the same as those introduced in the past studies but with different terminologies.
In the spherization radius introduced by Shakura \& Sunyaev (1973),
for example, $A_{\rm lau}$ was taken to be unity,
while $A_{\rm lau}\sim1.95$ in the critical radius introduced by Fukue (2004).
We wish to note that the launching radius is crudely equal to the photon-trapping radius.

\begin{table*}[h]
  \tbl{Results and initial settings of simulations}{
  \begin{tabular}{lllllllll}
    \hline
    paper & method & Compton & $r_{\rm out}$ & $r_{\rm K}$  & $r_{\rm qss}$ & $R_{\rm trap}$ & $\dot{M}_{\rm BH}$ & $\dot{M}_{\rm outflow}$ \\
     &             & [Yes/No]& $[r_{\rm S}]$ & $[r_{\rm S}]$& $[r_{\rm S}]$ & $[r_{\rm S}]$ & $[L_{\rm Edd}/c^{2}]$ & $[L_{\rm Edd}/c^{2}]$ \\
    \hline
    our simulation & 2D-RHD       & Yes  & $3000$ & $2430$& $\sim\rqss$  & $\sim\rtrap$   & $\sim\mdotinave$ &$\sim\mdotoutend$\\
    \hline
    Ohsuga$+$05    & 2D-RHD       & No  & $500$  & $100$ & $\sim30$   & $\sim200$  & $\sim130$ &\\
    Ohsuga$+$11    & 2D-RMHD      & No  & $105$  & $40$  & $\sim10$   & $\sim150$  & $\sim100$ &\\
    Jiang$+$14     & 3D-RMHD      & No  & $50$   & $25$  & $\sim20$   & $\sim330$  & $\sim220$ &$\sim400$\\
    S\c{a}dowski$+$15 & 2D-GR-RMHD & Yes & $2500$ & $21$  & $\sim35$   & $\sim640$  & $\sim420$ &$\sim7000$\\
    S\c{a}dowski$+$16 & 3D-GR-RMHD & Yes & $500$  & $20$  & $\sim10$   & $\sim260$  & $\sim180$ &$\sim520$\\
    Hashizume$+$15 & 2D-RHD       & No  & $5000$ & $100$  & $\sim100$ & $\sim230$  & $\sim150$ &$\sim500$\\
    Takahashi$+$16 & 3D-GR-RMHD    & No  & $125$  & $17$& $\sim10$   & $\sim300$  & $\sim200$ &\\
    Kitaki$+$18    & 2D-RHD       & Yes & $3000$ & $300$ & $\sim200$  & $\sim420$  & $\sim280$ &$\sim300$\\
    Jiang$+$19     & 3D-RMHD      & Yes & $800$  & $40$  & $\sim15$   & $\sim380$  & $\sim250$ &\\
    \hline
  \end{tabular}}
  \begin{tabnote}
    Here, $r_{\rm out}$ is the radius at the outer boundary,
    $r_{\rm K}$ is the initial Keplerian radius, 
    $r_{\rm qss}$ is the radius, inside which the quasi steady state is established,
    $R_{\rm trap}$ is the photon-trapping radius derived based on equation \ref{eq-trap},
    $\dot{M}_{\rm BH}$ is the accretion rate onto the black hole,
    and
    $\dot{M}_{\rm outflow}$ is the outflow rate at around $r_{\rm out}$.
    It is also indicated whether the Compton scattering effect is taken into account or not.
  \end{tabnote}
  \label{table1}
\end{table*}

Although the (semi-)analytical approaches are useful to understand the basics of the super-Eddington flow,
we need simulation studies, as well, to see what actually happens as the consequence of complex radiation-matter interactions.
Multi-dimensional RHD (radiation hydrodynamic) simulations of the super-Eddington accretion flow were pioneered by Eggum et al. (1988),
followed by Fujita \& Okuda (1998).
Their 2D (Two Dimensional) RHD simulations have shown that the super-Eddington accretion flow has a puffed-up structure and that high-speed outflow forms a funnel near the rotational axis.
Those simulation studies were, however, restricted within small computational boxes due to the regulation by the supercomputers available in those days.

More realistic and by far larger-scale simulation studies were initiated by Ohsuga et al. (2005), 
who performed much longer time-scale simulations and clarified the detailed properties of the accretion flow, outflow and the observational appearance of the super-Eddington systems.  
Since then rather extensive numerical simulation studies have been conducted;
first in Newtonian dynamics (e.g. Ohsuga et al. 2009, 2011; Kawashima et al. 2009; Jiang et al. 2014, 2019) 
and then in general relativistic treatment (e.g. McKinney et al. 2014; S\c{a}dowski et al. 2015, 2016; Takahashi et al. 2016).
In the rotating black hole, furthermore,
the emergence of strong and powerful jets driven by the Blandford-Znajek mechanism (Blandford and Znajek 1977) is expected,
which was calculated by GR-RMHD (general relativistic radiation magnetohydrodynamic) and GR-MHD (general relativistic magnetohydrodynamic) simulations.

We should note, however, that the authors adopted somewhat artificial numerical setting in these simulations for numerical reasons.
To be more precise, they started simulations by putting an initial torus near a black hole or
by injecting gas with small angular momentum so that a torus-like structure is formed near a black hole.

There are certainly cases, in which small $r_{\rm K}$ is expected, such as the case of tidal disruption events, but we focus on other cases with large $r_{\rm K}$, baring ULXs and NLS1s in mind.
Here, we define the Keplerian radius, $r_{\rm K}$, in such a way that the initial torus (with a given specific angular momentum) rotates around the central black hole with the Keplerian rotation velocity.
We also define the quasi-steady radius $r_{\rm qss}$, inside which quasi-steady state is achieved
(more rigorous definition will be given in section \ref{sec-mass-in-out}),
and list these values in table \ref{table1} for the recent simulation studies. 
From this table we understand that both of the Keplerian radius and quasi-steady radius are smaller than the trapping radius $R_{\rm trap}$ (equation \ref{eq-trap}) in all the past simulations.

Both were required for numerical reason, since otherwise it will take too long computational time to complete within a reasonable time, say, a few months.
But we should point that the previous simulation studies commonly exhibit a puffed up structure near the black hole, since $r_{\rm K}\ll R_{\rm trap}\sim R_{\rm lau}$,
and that a large amount of outflow material originates from such an inflated zone.
It may be possible that the outflow rate was grossly overestimated in such simulations (see table \ref{table1}).

In the present study, therefore,
we aim at expanding the quasi-steady region as much as possible so that it should cover the trapping radius and launching radius.
With this issue kept in mind, we perform 2D-RHD simulation of the super-Eddington accretion flow in a large calculation box, adopting a very large initial Keplerian radius (see, table \ref{table1}).
The main objectives of the present study are to clarify from which part of the accretion flow genuine outflow (that reaches the outer calculation boundary) is launched and how much mass, momentum, and mechanical energy is carried away by outflow in which direction. 
The plan of this paper is as follows:
We first explain our numerical methods and models in the next section.
We then present our results in section 3.
The discussion is given in section 4.
The final section is devoted to conclusions.

\section{Models and Numerical Methods}
\subsection{Radiation Hydrodynamic Simulations}
\label{sec-rhd}
In the present study, we consider super-Eddington accretion flow and outflow onto a black hole 
by injecting mass from the outer simulation boundary at a constant rate of $\dot{M}_{\rm input}$
with angular momentum.
The parameter values will be specified in section \ref{sec-initial-condition}.
The flux-limited diffusion approximation is adopted (Lervermore \& Pormaraning 1981; Turner \& Stone 2001).
We also adopt the $\alpha$ viscosity prescription (Shakura \& Sunyaev 1973).
General relativistic effects are incorporated by adopting the pseudo-Newtonian potential (Paczy\'{n}sky \& Wiita 1980).

Basic equations and numerical methods are the same as those in Kitaki et al. (2017, 2018),
but it is upgraded to solve the energy equations (see section \ref{sec-energy-solver}).
This 2D-RHD code solves the axisymmetric two-dimensional radiation hydrodynamic equations     
in the spherical coordinates $(x,y,z)=(r\sin\theta\cos\phi, r\sin\theta\sin\phi, r\cos\theta)$,
where the azimuthal angle $\phi$ is set to be constant.
We put a black hole with mass of $10 ~M_{\odot}$ at the origin.
In this paper, we distinguish $r$, radius in the spherical coordinates, and $R=\sqrt{x^{2}+y^{2}}$, radius in the cylindrical coordinates (e.g. equation \ref{eq-trap}).

The continuity equation is given by,
\begin{eqnarray}
  \frac{\partial \rho}{\partial t}+\nabla\cdot\left(\rho\bm{v}\right)=0.
\end{eqnarray}
Here, $\rho$ is the gas mass density and 
$\bm{v}=(v_{r},v_{\theta},v_{\phi})$ is the velocity of gas. 
Note that we retain the azimuthal component of the velocity.

The equations of motion are written as,
\begin{eqnarray}
  \frac{\partial (\rho v_{r})}{\partial t}+\nabla\cdot\left(\rho v_{r}\bm{v}\right)=-\frac{\partial p}{\partial r}+\rho\left(\frac{v_{\theta}^{2}}{r}+\frac{v_{\phi}^{2}}{r}-\frac{GM_{\rm BH}}{(r-r_{\rm s})^{2}}\right)\nonumber\\
  +\frac{\chi}{c}F_{0,r},\\
  \frac{\partial (\rho rv_{\theta})}{\partial t}+\nabla\cdot\left(\rho rv_{\theta}\bm{v}\right)=-\frac{\partial p}{\partial \theta}+\rho v_{\phi}^{2}\cot\theta\nonumber\\
  +r\frac{\chi}{c}F_{0,\theta},\\
  \frac{\partial (\rho r\sin\theta v_{\phi})}{\partial t}+\nabla\cdot\left(\rho r\sin\theta v_{\phi}\bm{v}\right)=\frac{1}{r^{2}}\frac{\partial}{\partial r}\left(r^{3}\sin\theta t_{r\phi}\right).
\end{eqnarray}
Here $p$ is the gas pressure,
$\chi=\kappa+\rho \sigma_{\rm T}/m_{\rm p}$ is the total opacity,
where $\kappa$ is free-free and free-bound absorption opacity
(Rybicki \& Lightman 1979),
$\sigma_{\rm T}$ is the cross-section of Thomson scattering,
$m_{\rm p}$ is the proton mass, and
$\bm{F}_{0}=(F_{0,r},F_{0,\theta},F_{0,\phi})$ is the radiative flux in the comoving frame,
where the suffix 0 represents quantities in the comoving frame and we set $F_{0,\phi}=0$.

We assume that only
the $r$-$\phi$ component of the viscous-shear tensor is nonzero,
and it is prescribed as
\begin{eqnarray}
  t_{r\phi}&=&\eta r \frac{\partial }{\partial r}\left(\frac{v_{\phi}}{r} \right),
\end{eqnarray}
with the dynamical viscous coefficient being
\begin{eqnarray}
  \eta&=&\alpha \frac{p+\lambda E_{0}}{\Omega_{\rm K}}.
\end{eqnarray}
Here,
$\alpha=0.1$ is the $\alpha$ parameter (Shakura \& Sunyaev 1973),
$\Omega_{\rm K}$ is the Keplerian angular speed,
$E_{0}$ is the radiation energy density,
and $\lambda$ represents the flux limiter of the flux-limited diffusion approximation
(Levermore \& Pormraning 1981; Turner \& Stone 2001).

The energy equations of gas and radiation are given by,
\begin{eqnarray}
  \frac{\partial e}{\partial t}+\nabla\cdot\left(e\bm{v}\right)&=&-p\nabla\cdot\bm{v}-4\pi\kappa B+c\kappa E_{0}\nonumber\\
  &&+\Phi_{\rm vis}-\Gamma_{\rm Comp},
\end{eqnarray}
and
\begin{eqnarray}
  \frac{\partial E_{0}}{\partial t}+\nabla\cdot\left(E_{0}\bm{v}\right)&=&-\nabla\cdot\bm{F}_{0}-\nabla\bm{v}:\bm{{\rm P}}_{0}+4\pi\kappa B-c\kappa E_{0}\nonumber\\
  &&+\Gamma_{\rm Comp},
\end{eqnarray}
respectively.
Here, $e$ is the internal energy density
which is linked to the thermal pressure by the ideal gas equation of state,
$p=(\gamma -1)e=\rho k_{\rm B}T_{\rm gas}/(\mu m_{\rm p})$
with $\gamma=5/3$ being the specific heat ratio,
$k_{\rm B}$ being the Boltzmann constant,
$\mu=0.5$ is the mean molecular weight (we assume pure hydrogen plasmas),
and $T_{\rm gas}$ is the gas temperature.
$B=\sigma_{\rm SB}T_{\rm gas}^{4}/\pi$ is the blackbody intensity
where $\sigma_{\rm SB}$ is the Stefan--Boltzmann constant.
$\bm{{\rm P}}_{0}$ is the radiation pressure tensor,
$\Phi_{\rm vis}$ is the viscous dissipative function written as
\begin{eqnarray}
  \Phi_{\rm vis}=\eta\left[r\frac{\partial }{\partial r}\left(\frac{v_{\phi}}{r} \right)\right]^{2}.
\end{eqnarray}
The Compton cooling/heating rate $\Gamma_{\rm Comp}$ is described as
\begin{eqnarray}
  \Gamma_{\rm Comp}&=&4\sigma_{\rm T}c\frac{k_{\rm B}\left(T_{\rm gas}-T_{\rm rad}\right)}{m_{\rm e}c^{2}}\left(\frac{\rho}{m_{\rm p}} \right)E_{0}.
\end{eqnarray}
Here, $m_{\rm e}$ is the electron mass and
$T_{\rm rad}\equiv (E_{0}/a)^{1/4}$ is the radiation temperature with the radiation constant $a=4\sigma_{\rm SB}/c$.

\subsection{Initial conditions and calculated models}
\label{sec-initial-condition}
Simulation settings are the also same as those in Kitaki et al. (2018) except for a larger value of $r_{\rm K}$ ($=2430~r_{\rm S}$).
The computational box is set by $r_{\rm in}=2~r_{\rm S}\leq r \leq r_{\rm out}=3000~r_{\rm S}$,    
and $0 \leq \theta \leq\pi/2$.
Grid points are uniformly distributed in logarithm in the radial direction;
$\triangle\log_{10} r = (\log_{10} r_{\rm out}-\log_{10} r_{\rm in})/N_{r}$,
while it is uniformly distributed in $\cos\theta$ in the polar direction;
$\triangle\cos \theta=1/N_{\theta}$, where the numbers of grid points     
are $(N_{r},N_{\theta})=(200,240)$.
We also simulated the case with $(N_{r},N_{\theta})=(400,480)$, confirming that our conclusions are not altered.

We initially put a hot optically thin atmosphere with negligible mass around the black hole for numerical reasons.
The initial atmosphere is assumed to be in isothermal hydrostatic equilibrium in the radial ($r$) direction.
Then, the density profile is
\begin{eqnarray}
  \rho_{\rm atm}(r,\theta)&\equiv&\rho_{\rm out}\exp\left[\frac{\mu m_{\rm p}GM_{\rm BH}}{k_{\rm B}T_{\rm atm} r_{\rm out}}\left(\frac{r_{\rm out}}{r}-1\right)\right],
\end{eqnarray}
where $\rho_{\rm out}$ is the density at the outer boundary.
We employ $\rho_{\rm out}=10^{-17}{\rm g~cm^{-3}}$ and $T_{\rm atm}=10^{11}{\rm K}$, following in Ohsuga et al. (2005).

Mass is injected continuously at a constant rate of $\dot{M}_{\rm input}$
through the outer disk boundary at $r=r_{\rm out}$ and $0.48\pi\leq\theta\leq0.5\pi$.
The black hole mass and mass injection rate are set to be
$M_{\rm BH}=10~M_{\odot}$ and $\dot{M}_{\rm input} = 700~L_{\rm Edd}/c^2$, respectively.
The injected gas is assumed to possess an specific angular momentum corresponding to the Keplerian radius of $r_{\rm K}=2430~r_{\rm S}$
(i.e., the initial specific angular momentum is $\sqrt{GM_{\rm BH} r_{\rm K}}$).
We thus expect that inflow material first falls towards the center
and forms a rotating gaseous ring at around $r\sim r_{\rm K}$, 
from which the material slowly accretes inward via viscous diffusion process.
We allow mass to go out freely through the outer boundary at $r=r_{\rm out}$
and $0\leq\theta\leq0.48\pi$.
and assume that mass at $r=r_{\rm in}$ is absorbed.

We assume that 
the density, gas pressure, radial velocity, and radiation energy density are symmetric at the rotational axis,
while $v_{\theta}$ and $v_{\phi}$ are the antisymmetric.
On the equatorial plane, on the other hand,
$\rho$, $p$, $v_{r}$, $v_{\phi}$ and $E_{0}$ are symmetric,
and $v_{\theta}$ is antisymmetric.
More details of the boundary conditions are written in Ohsuga et al. (2005).

\subsection{Updating energy equation solver}
\label{sec-energy-solver}
In the present study, we calculate a large scale structure of super-Eddington accretion flow
by adopting the total energy equation, instead of the internal energy equation.
This is preferable to calculate large-scale flow structure, 
since the total energy conservation does not hold in some cases, if we use the internal energy equation.

We employ the operator splitting method,
in which the viscous processes are separated from other processes (i.e., advection, radiation).
The viscosity-related terms in the equation of motion and energy equation are
\begin{eqnarray}
  \frac{\partial (\rho r \sin\theta v_{\phi})}{\partial t}&=&\frac{1}{r^{2}}\frac{\partial }{\partial r}\left(r^{3}\sin\theta t_{r\phi}\right), \label{eq-eom1}
\end{eqnarray}
and
\begin{eqnarray}
  \frac{\partial }{\partial t}\left(e + \frac{1}{2}\rho v_{\phi}^{2}\right)&=&\frac{1}{r^{2}}\frac{\partial }{\partial r}\left(r^{2}t_{r\phi}v_{\phi}\right),\label{eq-energy-vis-heat1}
\end{eqnarray}
respectively.
We solve these equations in the following way:
First, equation of motion (\ref{eq-eom1}) is solved by the implicit method through the Thomas algorithm (e.g. William et al. 2007),
and the velocity in the next time step $v_{\phi}^{n+1}$ is calculated.
Second, the total energy equation (\ref{eq-energy-vis-heat1}) is solved by using the velocity $v_{\phi}^{n+1}$, and the internal energy density in the next time step $e^{n+1}$ is obtained.
Then, other quantities will be updated.
The advantage of this method is that the total energy is always conserved.

\section{Results}

\subsection{Overall flow structure}
\label{sec-overall}
\begin{figure}[ht]
\begin{center}
  \includegraphics[width=80mm,bb=0 0 900 900]{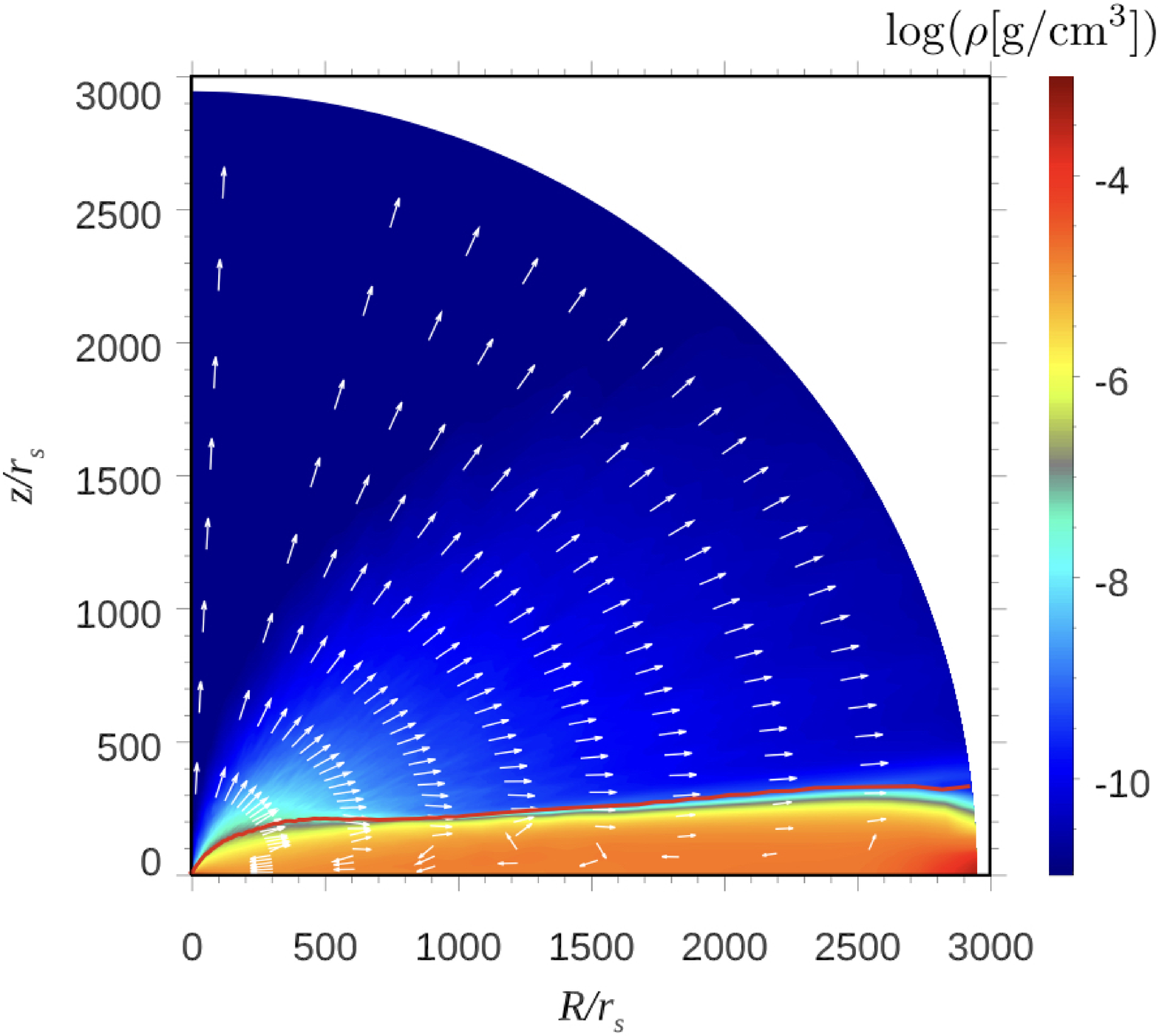}
\end{center}
\caption{
  Time-averaged density contours of super-Eddington accretion flow onto a black hole.
  Overlaid are the gas velocity vectors whose lengths are proportional to the logarithm of the absolute velocity.
  The red line represents the disk surface, which is defined as the loci where the radiation force balances the gravitational force.
}
\label{fig1}

\begin{center}
  \includegraphics[width=80mm,bb=0 0 900 900]{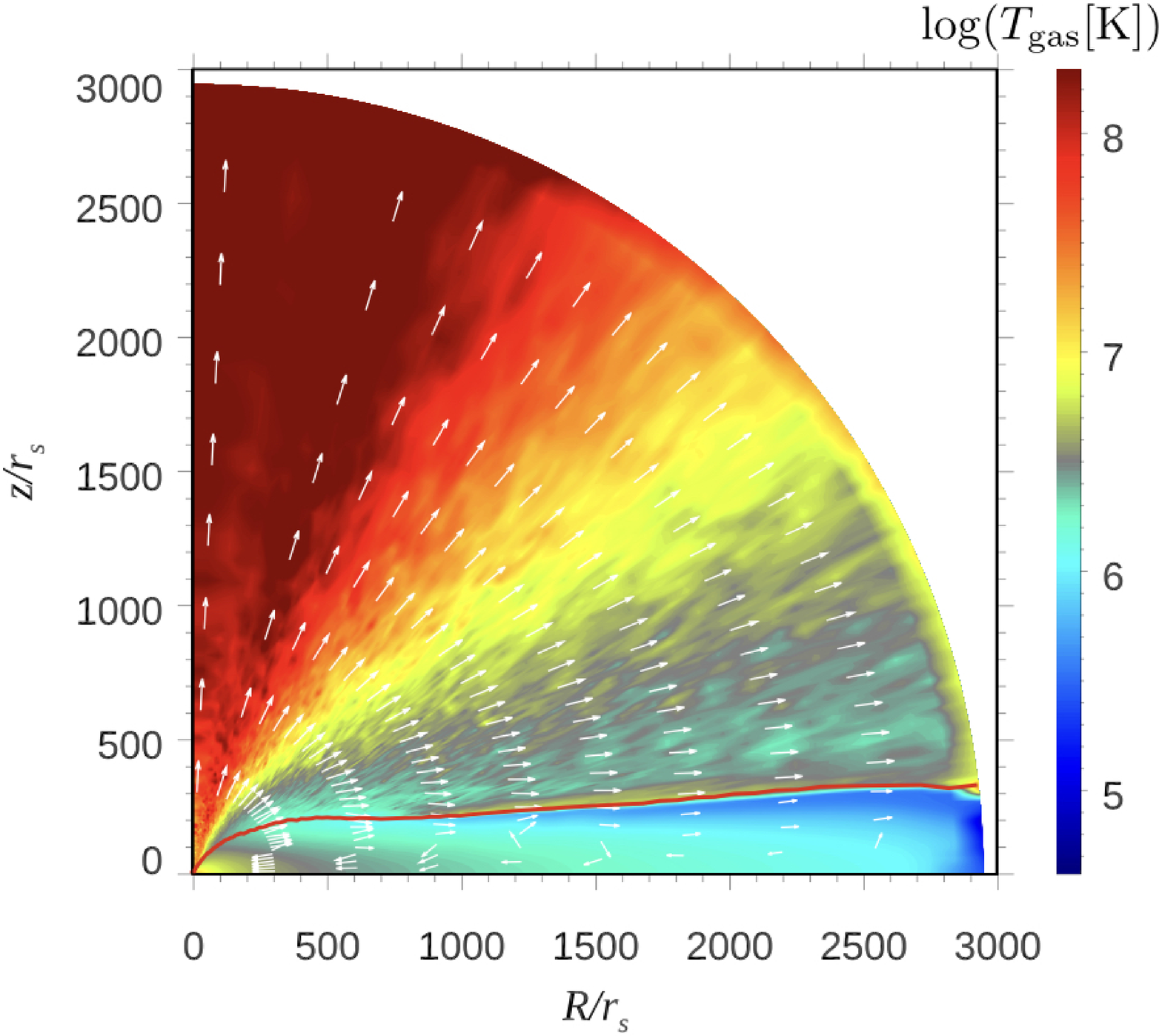}
\end{center}
\caption{Same as figure \ref{fig1} but for the temperature contours.
}
\label{fig2}
\end{figure}

\begin{figure}[ht]
\begin{center}
  \includegraphics[width=80mm,bb=0 0 900 900]{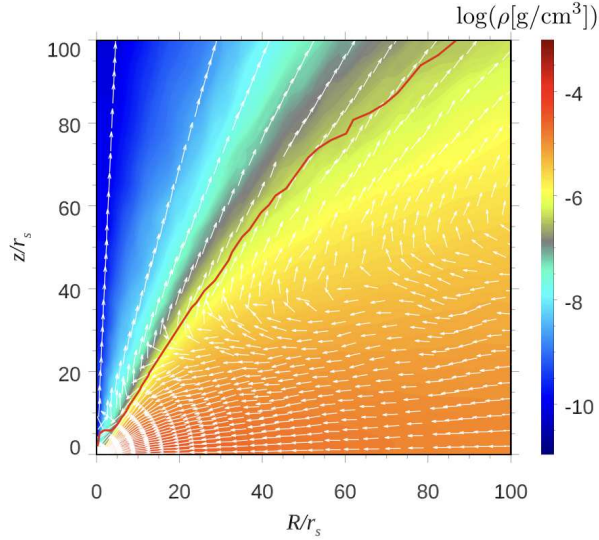}
\end{center}
\caption{
 Magnified view of the central region of figure \ref{fig1}.
}
\label{fig1_small}
\end{figure}

In this paper, we examine the time-averaged structure both of inflow and outflow in a quasi-steady state, unless stated otherwise.
We first show in figures \ref{fig1} and \ref{fig2} the density and temperature contours overlaid with the velocity fields in the quasi-steady state.
All the physical quantities (i.e., temperature, velocity, etc) except for gas mass density are time-averaged with weight of gas mass density during the interval of $t\sim8519$--$9109~{\rm sec}$, while gas mass density is simple time averages with no weight.
After the simulation starts,
the gas injected from the outer boundary into an initially empty zone
first free-falls and accumulates around the initial Keplerian radius,
$r_{\rm K} \sim 2430 ~r_{\rm S}$,
since the centrifugal force and the gravitational force balance there.
Soon after the transient initial phase accumulated matter spreads outward and inward in the radial direction via viscous diffusion process,
forming an accretion disk extending down to the innermost zone ($t \lesssim 8511~{\rm sec}$).
The newly injected matter collides with the disk matter so that a high-density region appears at $\sim(2400$--$3000)r_{\rm S}$ 
(well outside the initial Keplerian radius) in figure \ref{fig1}. 
In a sufficiently long time (on the order of the viscous timescale, $t\gtrsim8511~{\rm sec}$, Ohsuga et al. 2005),
quasi-steady, inflow-outflow structure is established (see figure \ref{fig1}).

In figures \ref{fig1} and \ref{fig2} we also indicate the disk surface by the red solid line.
By the disk surface we mean the loci, in which radiation force balances the gravity in radial direction,
$\chi F_{0,r}/c=\rho GM_{\rm BH}/(r-r_{\rm S})^{2}$.
We examined other definitions of the disk surface; for example,
we do the same but including the centrifugal force term, $\rho v_{\phi}^{2}/r$
but found no better definitions in a simple form.

We notice a quite different flow shape in figure \ref{fig1}
from those reported previously (see table \ref{table1}).
That is, we no longer find a puffed-up structure, which was commonly observed in the previous studies, but find a rather smooth disk shape up to the outer boundary(see model a11 of figure 1 in Kitaki et al. 2018). 
This is because we adopted a very larger Keplerian radius than that of the previous simulations.
The disk height is roughly proportional to $R$ in the inner region, $R \lesssim 300 ~r_{\rm S}$, 
whereas it is roughly constant outside ($H\sim200~r_{\rm S}$--$300~r_{\rm S}$).

We also plot the velocity fields of gas by the white vectors in figures \ref{fig1} and \ref{fig2}.
We understand that gas is stripped off from the disk surface to form outflow.
Near the rotation axis, especially,
we see a cone-shaped funnel filled with high velocity ($\sim 0.3~c$) and high temperature plasmas of $T_{\rm gas}\gtrsim 10^{8}~{\rm K}$,
surrounded by the outflow region of modest velocity ($\sim 0.05~c$--$0.1~c$) and modest temperatures,
$T_{\rm gas}\sim10^{6-7}~{\rm K}$.

Figure \ref{fig1_small} is the magnification of the central region of figure \ref{fig1}.
When we have a closer look at the disk region, we notice that 
the gas motion is outward near the surface, whereas it is inward near the equatorial plane.
These velocities reflect convective motion (to be discussed later in section \ref{sec-convection}).
We also notice that almost all the velocity vectors (except for some near the disk surface) 
are inward (i.e., towards the central black hole) at $r \lesssim \rinflow r_{\rm S}$,
although a hint of convective motions is observed in snapshots (to be discussed in section \ref{sec-convection}).

\subsection{Mass inflow rate and mass outflow rate}
\label{sec-mass-in-out}

\begin{figure}[ht]
\begin{center}
  \includegraphics[width=80mm,bb=0 0 360 288]{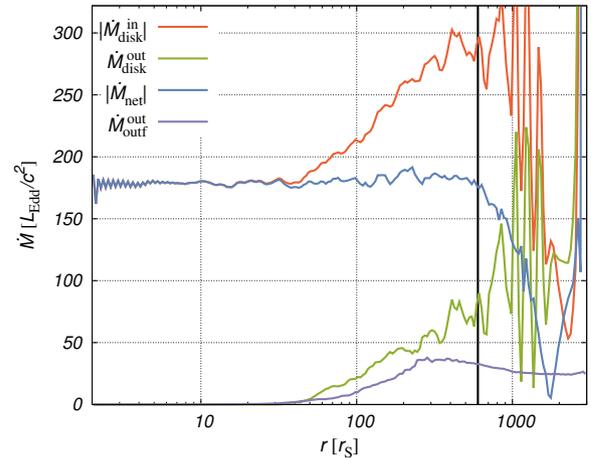}
\end{center}
\caption{
  Time-averaged radial profiles of the mass inflow rate, $\dot{M}_{\rm disk}^{\rm in}$ (red line),
  the mass outflow rate both within the disk, $\dot{M}_{\rm disk}^{\rm out}$ (green line),
  the net flow rate, $\dot{M}_{\rm net}$ (blue line),
  and the mass outflow rate in the outflow region (above the disk surface), $\dot{M}_{\rm outf}^{\rm out}$ (purple line), respectively.
  The net flow rate is nearly constant inside the quasi-steady radius, 
  $r_{\rm qss}\sim\rqss ~r_{\rm S}$, which is indicated by the vertical black line.
}
\label{fig3}
\end{figure}

The radial profiles of the mass flow rates are a very useful tool to diagnose gas dynamics around the black hole.
We calculate the five flow rates:
the mass inflow and outflow rates in the disk,
the same in the outflow region (the region above the disk surface),
and the net flow rate:
\begin{eqnarray}
  \dot{M}_{\rm disk}^{\rm in}(r)&\equiv&4\pi\int_{\theta_{\rm surf}}^{\pi/2} d\theta\sin\theta\nonumber\\
  &&~~~~~~~~~\times r^{2}\rho(r,\theta){\rm min}\left\{v_{r}(r,\theta),0\right\},\\
  \dot{M}_{\rm disk}^{\rm out}(r)&\equiv&4\pi\int_{\theta_{\rm surf}}^{\pi/2} d\theta\sin\theta\nonumber\\
  &&~~~~~~~~~\times r^{2}\rho(r,\theta){\rm max}\left\{v_{r}(r,\theta),0\right\},\\
  \dot{M}_{\rm outf}^{\rm in}(r)&\equiv&4\pi\int_{0}^{\theta_{\rm surf}} d\theta\sin\theta\nonumber\\
  &&~~~~~~~~~\times r^{2}\rho(r,\theta){\rm min}\left\{v_{r}(r,\theta),0\right\},   \label{eq-mdotinflow}\\
  \dot{M}_{\rm outf}^{\rm out}(r)&\equiv&4\pi\int_{0}^{\theta_{\rm surf}} d\theta\sin\theta\nonumber\\
  &&~~~~~~~~~\times r^{2}\rho(r,\theta){\rm max}\left\{v_{r}(r,\theta),0\right\},   \label{eq-mdotoutflow}\\
  \dot{M}_{\rm net}(r)&\equiv&\dot{M}_{\rm disk}^{\rm in}(r)+\dot{M}_{\rm disk}^{\rm out}(r) + \dot{M}_{\rm outf}^{\rm in}(r) + \dot{M}_{\rm outf}^{\rm out}(r).
\end{eqnarray}
Here, $\theta_{\rm surf}=\theta_{\rm surf}(r)$ is the angle between the rotation axis and the disk surface (see the red line in figure \ref{fig1}).

Figure \ref{fig3} illustrates the absolute values of the various mass flow rates as functions of radius, $r$,
except for $\dot{M}_{\rm outf}^{\rm in}$, since it turns out to be practically zero.

Let us first focus on the blue line which stands for the net accretion rate,
since this line provides a key information to evaluate to what extent a quasi-steady state is achieved.
We see that this line is approximately constant in the range of $r=2~r_{\rm S}$--$\rqss ~r_{\rm S}$;
that is $r_{\rm qss}\equiv \rqss ~r_{\rm S}$.
This value is unprecedently large (see table \ref{table1}), and, hence,
the present simulation can provide us with much more reliable information on the outflow properties.

Let us next examine the behavior of the various lines in the innermost region ($r < r_{\rm inflow}\sim40~r_{\rm S}$).
We find negligibly small mass outflow rate (both of $\dot{M}_{\rm disk}^{\rm out}$, 
$\dot{M}_{\rm outf}^{\rm out}$), while the mass inflow rate stays constant.
This feature agrees well with that of our previous calculations (Kitaki et al. 2018).
The mass inflow and outflow rates averaged over the range of $r=2~r_{\rm S}$--$30~r_{\rm S}$ are
$\dot{M}_{\rm BH}\equiv\langle|\dot{M}_{\rm disk}^{\rm in}|\rangle=\mdotinave~L_{\rm Edd}/c^{2}$,
$\langle\dot{M}_{\rm disk}^{\rm out}\rangle=\mdotoutavedisk~L_{\rm Edd}/c^{2}$
and
$\langle\dot{M}_{\rm outf}^{\rm out}\rangle=\mdotoutave~L_{\rm Edd}/c^{2}$,
respectively.

One may think that such a negligibly small mass outflow rate from the innermost region seems to be against a naive expectation that the smaller the radius is, the large becomes radiation-pressure force and so does the mass outflow rate.
This is not the case, however, since the radiation flux in the co-moving frame is rather inward in the innermost region because of photon trapping (see section \ref{sec-r-trap}).
In other words, the density on the disk surface
(where the radiation-pressure force balances with the gravitational force, see section \ref{sec-overall}) 
decreases inward so that the outflow rate ($\propto\rho v_{r}$) should also decrease (see Kitaki et al. 2018).

We are now ready to examine from which part of the accretion flow outflow emerges by the examination of the lines in the middle region ($40~r_{\rm S}$ -- $600 ~r_{\rm S}$).
The outflow rate above the disk surface ($\dot{M}_{\rm outf}^{\rm out}$),
indicated by the purple line in figure \ref{fig3}, increases with increasing radius,
reaches its maximum value of $38~L_{\rm Edd}/c^2$ at $r=\rlau ~r_{\rm S}(\equiv r_{\rm lau})$,
and then decreases beyond.
How can we understand this?

Here, we wish to stress that the outflow rate ($\dot{M}_{\rm outf}^{\rm out}$) plotted in figure \ref{fig3} is the cumulative one.
To be precise, $\dot{M}_{\rm outf}^{\rm out}(r)$ is defined by the mass flow rate measured at the radius $r$ and so we take into account all the materials which pass through the shell at $r$ above the disk surface, irrespective the launching points (see equation \ref{eq-mdotoutflow}).
Therefore, 
the outflow rate should monotonically increase with increase of the radius, 
as long as outflow occurs.
At even larger radii ($r>\rlau ~r_{\rm S}$), however,
$\dot{M}_{\rm outf}^{\rm out}$ decreases with increase of radius.
This is because the outflowing gas which was launched at smaller radii partly 
goes back to the disk surface. 
We call this sort of outflow as {\lq\lq}failed outflow{\rq\rq} 
(see next subsection for further discussion).
Note that the genuine outflow rate (by the purple line) 
is entirely less than the outflow rate in the disk (by the green line),
$\dot{M}_{\rm disk}^{\rm out}$,
which is caused by radial convective motion within the disk.

In the further outer region, $r\gtrsim 1000~r_{\rm s}$, $\dot{M}_{\rm outf}^{\rm out}$ is nearly constant.
The space-averaged (genuine) outflow rate at $r=2000~r_{\rm S}$--$3000~r_{\rm S}$ is 
$\dot{M}_{\rm outflow}\equiv\langle\dot{M}_{\rm outf}^{\rm out}\rangle\sim \mdotoutend ~L_{\rm Edd}/c^{2}$.

\subsection{Outflow streamlines}
\label{sec-streamline}
\begin{figure}[h]
\begin{center}
  \includegraphics[width=80mm,bb=0 0 900 1100]{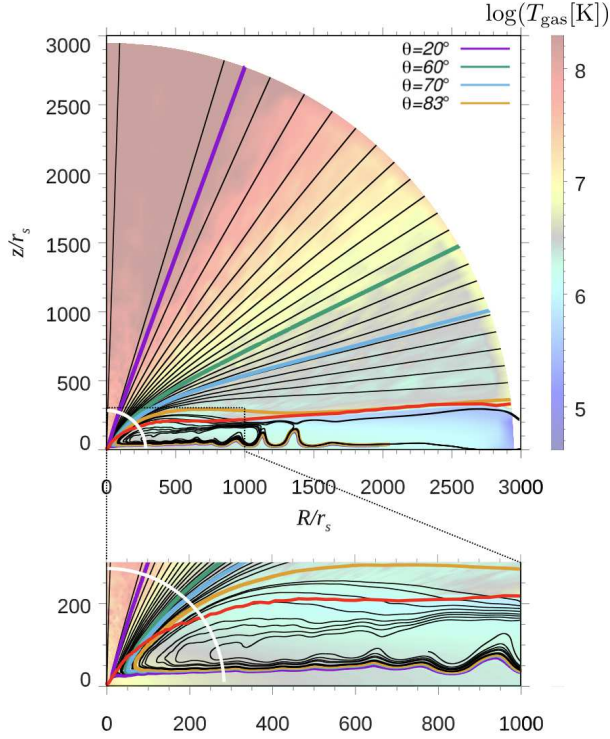}
\end{center}
\caption{
  Sequence of the streamlines overlaid on the gas temperature contours.
  The upper panel is the large-scale view, while the lower panel is the magnification of the central region.
  In each panel we pick up several streamlines and colored them:
  the purple, green, light blue, and orange lines, respectively, represent
  the streamlines which approaches the lines of constant 
  $\theta = 20^{\circ}$, $60^{\circ}$, $70^{\circ}$, and $83^{\circ}$ at $r=3000~r_{\rm S}$, respectively.
  The red line indicates the disk surface, and the white line represents the loci of 
  $r_{\rm lau}=\rlau ~r_{\rm S}$, at which the cumulative outflow rate reaches its maximum.
}
\label{fig6}
\end{figure}

\begin{figure}[h]
\begin{center}
  \includegraphics[width=80mm,bb=0 0 360 648]{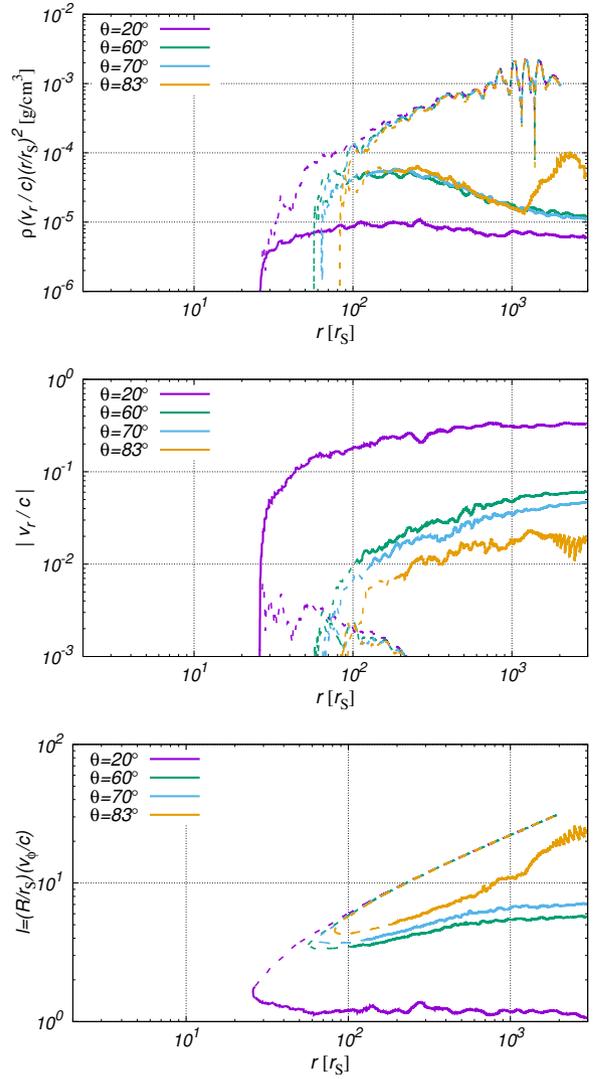}
\end{center}
\caption{
  Variations of some physical quantities along the four colored streamlines shown in figure \ref{fig6}:
  the mass flow rates per unit solid angle (top),
  the absolute values of the radial velocity (middle),
  and the specific angular momenta (bottom). 
  The solid lines (or the dashed lines) indicate the quantities in the outflow (disk) region.
}
\label{fig9}
\end{figure}

The streamline analysis is a powerful tool to investigate the physical evolution of the outflowing gas after being launched from a certain point on the disk surface.
Figure \ref{fig6} displays a sequence of streamlines overlaid on the temperature contours.
(The lower panel is the magnified view of the central region of the upper panel.)
We understand in this figure that the outflow emerges from the disk surface inside the white circle put at radius, $r = \rlau ~r_{\rm S}$,
where $\dot{M}_{\rm outf}^{\rm out}$ reaches its maximum (see figure \ref{fig3}).
We, here, define the launching radius ($R_{\rm lau}$) to be the radius,
where the white line crosses the red line; that is,
\begin{eqnarray}
  R_{\rm lau}&=&\rlau ~r_{\rm S}\sin(55^{\circ})\sim\rlaucyl~r_{\rm S}
\sim1.3~\dot{m}_{\rm BH}r_{\rm S}.
\end{eqnarray}
This value is consistent with the analytical estimation (equation \ref{eq-launching}).

The streamlines between the orange and red lines in figure \ref{fig6} 
represent the failed outflow;
that is, the outflow which once leaves the disk surface at smaller radii but eventually comes back to the disk at large radii.
The launching radius of the genuine outflow (which can reach the outer boundary of the computational box) is given by,
\begin{eqnarray}
  R_{\rm lau}^{\infty}&=&\rlauinf~r_{\rm S}\sin(46^{\circ})
\sim\rlauinfcyl~r_{\rm S}\sim0.75~\dot{m}_{\rm BH}r_{\rm S}. 
\end{eqnarray}

The Bernoulli parameter, $Be$, is everywhere negative in the failed outflow region,
so is the pure outflow region around the disk surface.
We should note, however, that the Bernoulli parameter is not a conserved quantity in the viscous flow,
and that $Be$ is positive far from the black hole in the pure outflow region.

It is curious to examine how physical quantities vary along each streamline.
Figure \ref{fig9} illustrates the variations of the physical quantities along each of the
colored streamlines depicted in figure \ref{fig6}.
Let us first see the purple solid line 
(this streamline is connected to the funnel region, see figure \ref{fig6}).
From the middle panel we understand that the gas is quickly accelerated to finally acquire a high velocity, $v_{r}\sim 0.3~c$, within the funnel.
The outflow rate ($r^{2}\rho v_{r}\sim {\rm const}$) is conserved along the streamline 
within the funnel (see the upper panel).
Thus, the radial profile of the gas mass density is roughly $\rho \propto r^{-2}$.
The specific angular momentum ($R v_{\phi}$) is also conserved (see the lower panel).
This can be easily understood, since the viscosity is not effective in the outflow region.

Let us next consider the green and light blue lines, both of which are connected to
the (genuine) outflow region in figure \ref{fig9}.
From the middle panel, we see that the radial velocities are gradually accelerated 
until they reach the final value of several tenths of $c$ at around $r\gtrsim10^{3}~r_{\rm S}$.
Again, the specific angular momenta are roughly conserved, as is shown in the bottom panel.

The orange solid line in each panel of figure \ref{fig9} shows similar tendencies to those of the green and light blue solid lines, except in the region around $r\sim 3000~r_{\rm S}$,
where the outflowing gas comes back to the disk surface region and merges there.

\subsection{photon-trapping radius}
\label{sec-r-trap}
\begin{figure}[h]
\begin{center}
  \includegraphics[width=80mm,bb=0 0 360 288]{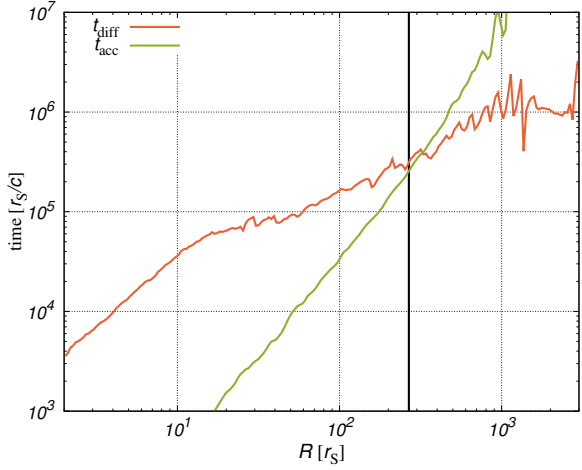}
\end{center}
\caption{
  The radiative diffusion timescale (red) and the dynamical timescale (green) both evaluated on the equatorial plane.
  The vertical black line is the photon-trapping radius calculated based on the slim disk model ($R_{\rm trap}\sim\rtrap ~r_{\rm S}$).
}
\label{fig11}
\end{figure}

The most principal value characterizing the super-Eddington accretion disk is the photon-trapping radius (see equation \ref{eq-trap}).
The photon-trapping radius is the radius where the radiative diffusion timescale $t_{\rm diff}$ is equal to the dynamical timescale $t_{\rm dyn}$.
\begin{eqnarray}
  t_{\rm diff}&=&\frac{H_{\rm inf}(R)}{c/[3\tau_{\rm e}(R)]},\\
  t_{\rm dyn}&=&\frac{R}{|v_{r}(R,z=0)|}.
\end{eqnarray}
Here,
$H_{\rm inf}$ is the height under which the radial velocity is negative,
and
$\tau_{\rm e}$ is vertical Thomson optical depth measured from equatorial plane to $z=H_{\rm inf}(R)$.
Figure \ref{fig11} shows these two timescales and they intersect each other at $R_{\rm trap}\sim\rtrapsph~r_{\rm S}$.
By contrast, the photon-trapping radius which the slim disk model predict is $R_{\rm trap}\sim \rtrap ~r_{\rm S}$ (see equation \ref{eq-trap}).
This value is close the one estimated numerically in the present study.

\begin{figure}[ht]
\begin{center}
  \includegraphics[width=80mm,bb=0 0 360 288]{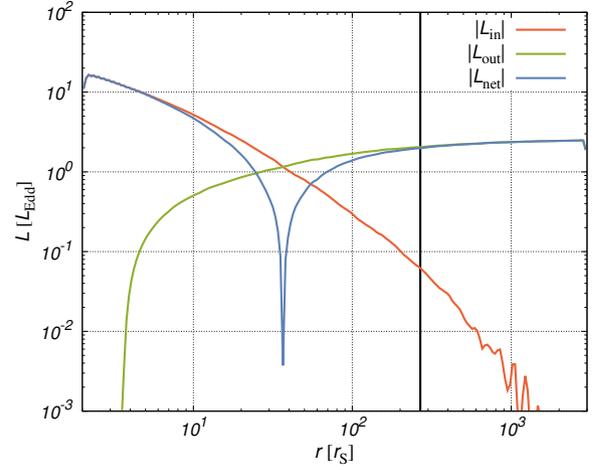}
\end{center}
\caption{
  The radial profiles of the inward, outward and net luminosities.
  The vertical black line indicates the photon-trapping radius based on the slim disk model
  ($R_{\rm trap}\sim\rtrap ~r_{\rm S}$).
}
\label{fig12}
\end{figure}

We evaluate the photon-trapping radius from another viewpoint.
The inward, outward and net luminosities are, respectively, written as,
\begin{eqnarray}
  L_{\rm in}(r)&=&\int_{4\pi}d\Omega~ r^{2}{\rm min}\left\{F_{\rm lab}^{r},0\right\},\\
  L_{\rm out}(r)&=&\int_{4\pi}d\Omega~ r^{2}{\rm max}\left\{F_{\rm lab}^{r},0\right\}, \label{eq-outlum}
\end{eqnarray}
and
\begin{eqnarray}
  L_{\rm net}(r)&=&L_{\rm in}(r)+L_{\rm out}(r).
\end{eqnarray}
Here, $F_{\rm lab}^{r}$ is the radial component of radiation flux in the laboratory frame.
In figure \ref{fig12} we compare these luminosities.
The inward luminosity, which represents the photon-trapping effect, increases inward
and the two lines intersect at $r \sim 38~r_{\rm S}$,
which is significantly less than the photon-trapping radius derived based on the slim disk model; i.e., $R_{\rm trap}\sim \rtrap ~r_{\rm S}$.
If we take the radius, where $ L_{\rm in}(r)$ vanishes, it is $\sim 10^{3} ~r_{\rm S}$,
much larger than $R_{\rm trap}$.

\begin{figure}[ht]
\begin{center}
  \includegraphics[width=80mm,bb=0 0 800 400]{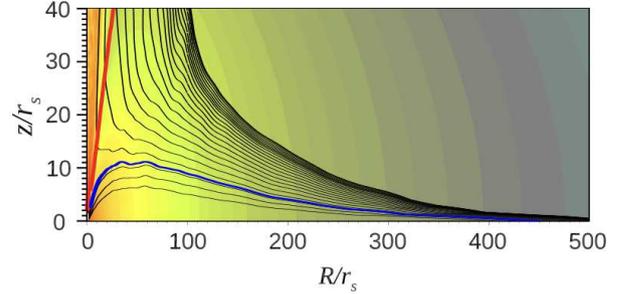}
\end{center}
\caption{
  The radiation flux-line (black lines) overlaid on the gas temperature contours (figure \ref{fig2}).
  The red line indicates the disk surface.
  The blue line starts at $R\sim\rtrapf~r_{\rm S}$ on the equatorial plane is connected to the black hole because of the photon-trapping effect.
}
\label{fig18}
\end{figure}

Why are these estimations so distinct?
We should note here that the diffusion timescale depends on the vertical position of the region in question;
the larger the vertical position ($z$) is, the shorter becomes the diffusion timescale.
We thus made another analysis; namely, we continuously connect the radiation flux vectors
starting from the disk surface, as we did in the streamline analysis to follow gas motion, 
and display the resultant {\lq\lq}flux-line{\rq\rq} in figure \ref{fig18}.
(Note that this flux-line does not have the same meaning of streamlines, since divergence of the radiation flux is not zero and since photons diffuse way as they proceed inward.) 
We see that the flux-lines starting from the disk surface at $R < \rtrapf~r_{\rm S}\sim 2.5~\dot{m}_{\rm BH}r_{\rm S}$
are finally connected to the black hole region.
We thus take this value as the numerical trapping radius, which is close to the analytical estimation.

The previous GR-RMHD simulations revealed that the magnetic fields help the photons to escape from the disk surface,
since the gas moves towards the disk surface by the magnetic buoyancy
and the photons are trapped in the gas (e.g., Blaes et al. 2011; Jiang et al. 2019).
Therefore, the photon-trapping radius may become smaller in the GR-RMHD simulations.

\subsection{Convection in the accretion disk}
\label{sec-convection}

\begin{figure}[h]
\begin{center}
  \includegraphics[width=80mm,bb=0 0 800 800]{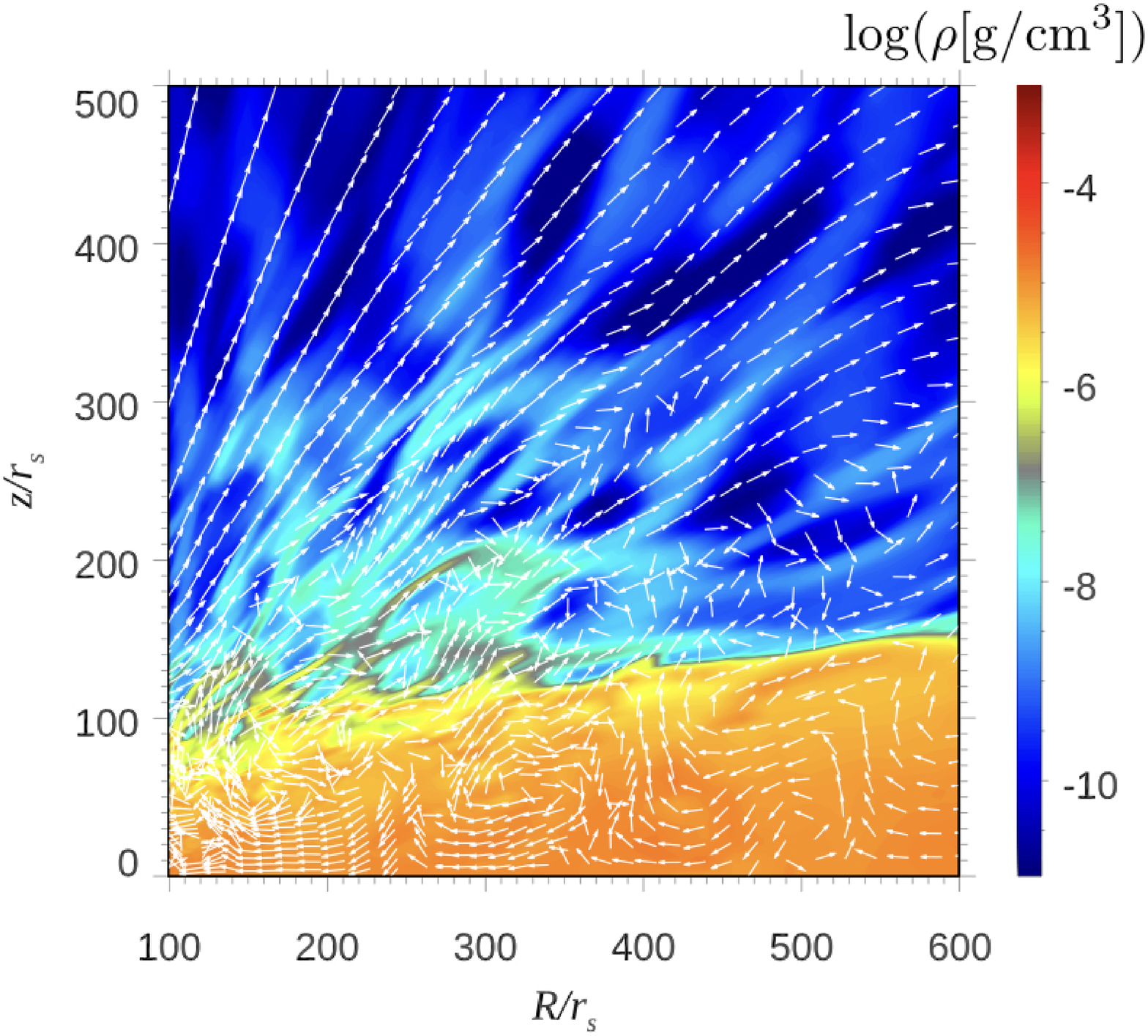}
\end{center}
\caption{
  The snapshot of the density contours overlaid with velocity vectors in the region
  between $100~r_{\rm S}\leq R \leq 600~r_{\rm S}$.
  We clearly see convection cells in the disk.
}
\label{fig13}

\begin{center}
  \includegraphics[width=80mm,bb=0 0 800 800]{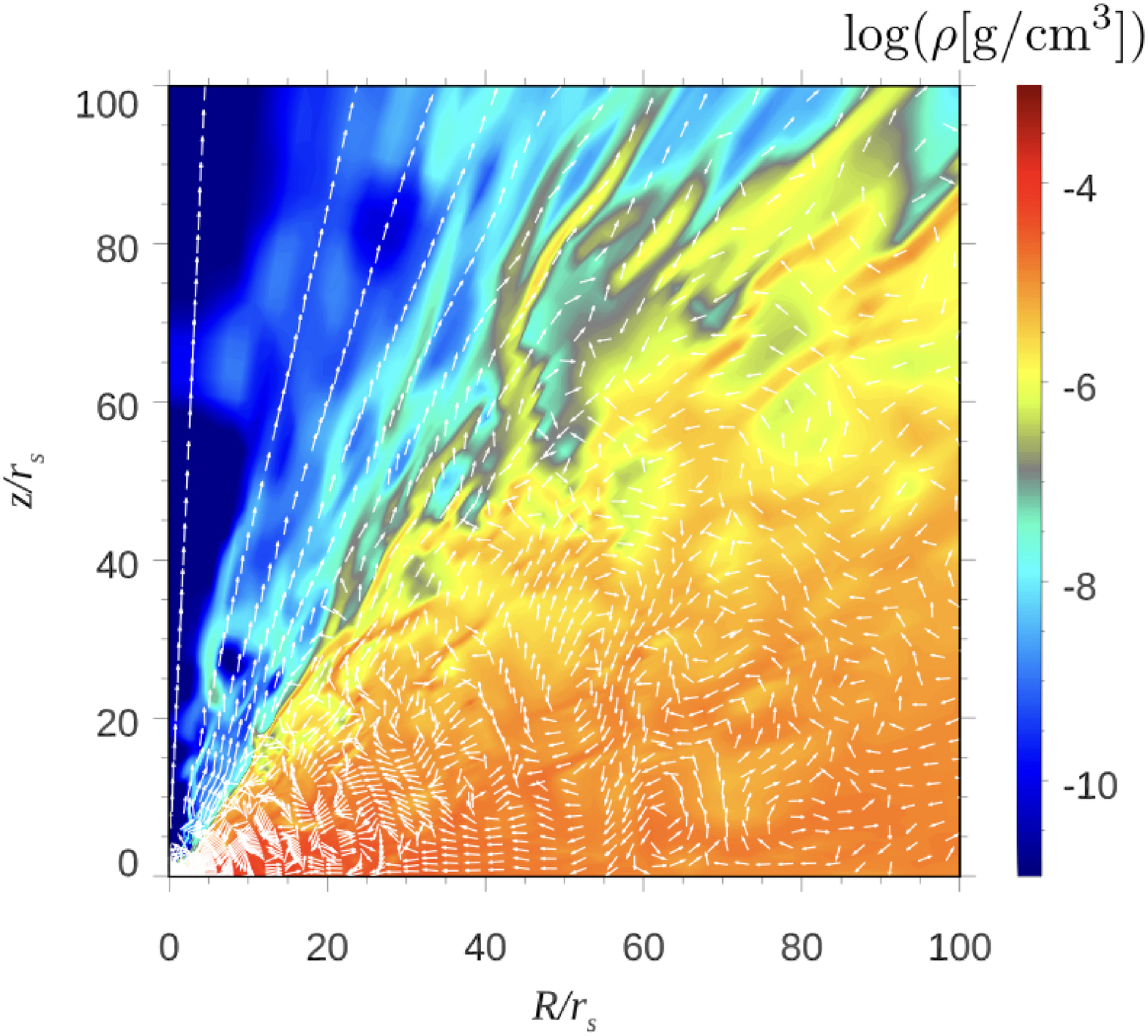}
\end{center}
\caption{
  Same as figure \ref{fig13} but in the central region near the black hole.
  We see convective cells even in the vicinity of the black hole at $r < r_{\rm inflow}\sim40~r_{\rm S}$, 
  which were not visible in the time-averaged profile (see figure \ref{fig1_small}).
}
\label{fig113}
\end{figure}

It is well known that the super-Eddington accretion flow undergoes large-scale circular gas motion or convection (see, e.g., Eggum et al. 1988).
Since the entropy generated within the accretion disk cannot easily be taken away out of 
the disk because of the inefficient radiative diffusion in the vertical direction,
the entropy tends to increase as gas accretes, condition for convective instability, 
as in the case of radiatively inefficient flow (see, e.g., Narayan \& Yi 1994).

Pietrini et al. (2000) proved that the convection which the vertical wavelength is larger than the radial wavelength occurs in the radiation-pressure dominated disk 
by deriving the dispersion relation from the RHD equations.
Some analytical and semi-analytical studies were made to describe the vertical structure 
under the assumption of separating the radial and vertical structures.
For example, S\c{a}dowski et al. (2011) calculated the vertical structure of super-Eddington accretion disk by the Runge-Kutta method, and showed that the energy is transported in vertical direction by the convection.
While, Gu (2012) derived the self-similar solution of super-Eddington accretion disk by assuming the radial dependence of the gas mass density and the radiation energy density, and this solution does not satisfy the convection criterion.
Thus, the occurrence of convection is a controversial issue and the previous studies
may depend on the various assumptions for the way of splitting the radial and vertical structure.

To proceed, it is useful to perform numerical simulations.
It seems important to note that the convection also occurs in the inner region of the standard disk where radiation pressure is dominant.
This is because the radiation diffusion becomes inefficient as in the case of the super-Eddington accretion flow.
According to the RHD simulations of the standard disk, the energy in the radiation pressure dominated region is transported not by the radiation diffusion but by the advection of the radiation (Agol et al. 2001, Blaes et al. 2011).
Considering these studies about the standard disk,
we conclude that analyzing the energy budget is very useful to evaluate the convection in the super-Eddington accretion disk.

As mentioned in section \ref{sec-mass-in-out},
convection occurs in the super-Eddington accretion disk.
Figures \ref{fig13} and \ref{fig113} show the snapshots of the cross-sectional view of the accretion flow.
We see there circular motions of velocity vectors around several points, e.g.,
$(R,~z)=(17~r_{\rm S},~10~r_{\rm S})$, $(50~r_{\rm S}, ~20~r_{\rm S})$, $(320~r_{\rm S},~50~r_{\rm S})$, and $(470~r_{\rm S}, ~110~r_{\rm S})$.
These convections tend to rotate in the clockwise direction.
This is because the radiation force and the centrifugal force tend to overcome the gravitational force when the gas blob rises from the equatorial plane to the surface of the disk.

It is very important, however, to note that small-scale convective motions totally disappear and the global convection appear when we make time average (see figures \ref{fig1} and \ref{fig2}).
Here, we wish to emphasize that this global convective motion is constructed by time-averaging the small-scale circular motions.
The time-averaged direction of the gas motion is inward near the equatorial plane,
while it is the outward near the surface of the accretion disk.
The inflow and outflow motions in the disk by the global convection are dominant in the entire accretion flow
(see the red and green lines in comparison with purple line in figure \ref{fig3}).

It is previously indicated that the occurrence of convection in snapshots may depend on the adopted $\alpha$-parameter;
the smaller $\alpha$ is, more efficient becomes convection motion (Igumenshchev et al. 1999, 2000; Yang et al. 2014),
and the $\alpha$-parameter decreases with increasing radius ($\alpha\sim 0.05$--$0.2$, Jiang et al. 2019).
Therefore, the (time-averaged) global convection may be modified
if we adopt smaller $\alpha$-parameter values and/or if we perform RMHD simulations.

\subsection{Energy transportation by convective motion}
\label{sec-convection2}

\begin{figure}[htb]
\begin{center}
  \includegraphics[width=80mm,bb=0 0 360 648]{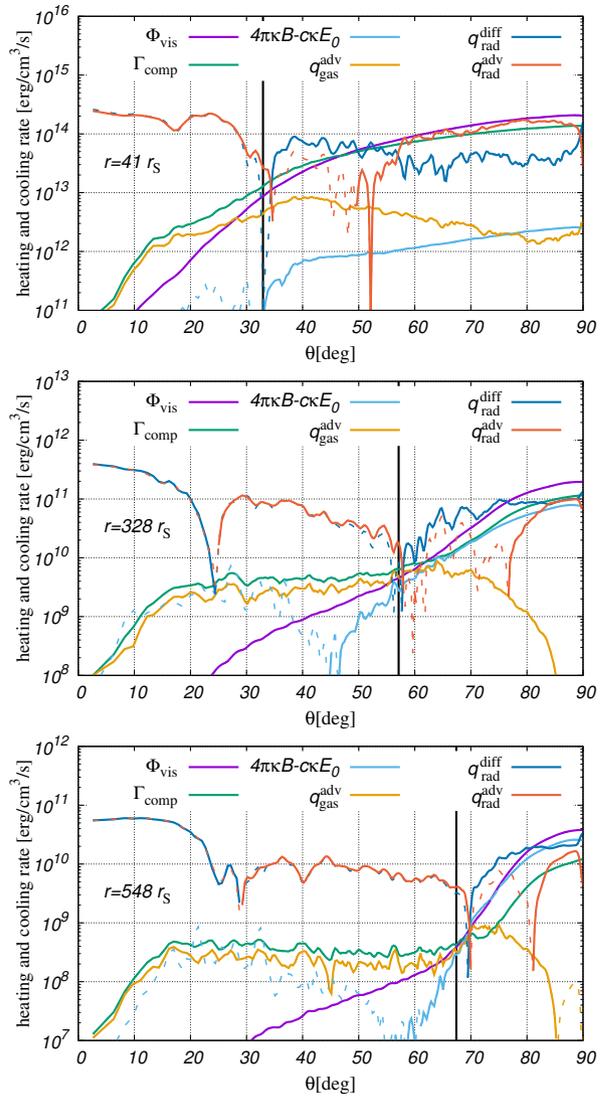}
\end{center}
\caption{
  Polar angle dependences of the heating and cooling rates at three different radii:
  $r=41~r_{\rm S}$ (top), $r=328~r_{\rm S}$ (middle), and $r=548~r_{\rm S}$ (bottom).
  The colored lines represent the following quantities:
  the viscous heating (purple),
  the net radiation heating and cooling (light blue),
  the advection of the radiation (red),
  the Compton heating and cooling (green),
  the advection of the gas (orange),
  and the radiation diffusion (blue), respectively.
  The black vertical line indicates the polar angle of the disk surface (see figure \ref{fig1}).
  The solid lines mean that each value is positive (e.g. $4\pi\kappa B -c\kappa E_{0} >0$),
  whereas the dashed lines mean that each value is negative (e.g. $4\pi\kappa B -c\kappa E_{0} <0$).
}
\label{fig17}
\end{figure}

Plotting the heating and cooling rates is a powerful tool to diagnose gas dynamics in the disk, as was demonstrated by Blaes et al. (2011).
In the gas energy equation,
viscous heating and advective heating balance with the energy transported to the radiation,
\begin{eqnarray}
  \Phi_{\rm vis} +q_{\rm gas}^{\rm adv} &=& \Gamma_{\rm comp} + (4\pi\kappa B -c\kappa E_{0}), \label{eq-HC0}
\end{eqnarray}
while in the radiation energy equation,
energy transported from the gas balances with the radiation advection cooling and radiative diffusion,
\begin{eqnarray}
  \Gamma_{\rm comp} + (4\pi\kappa B -c\kappa E_{0})&=& q_{\rm rad}^{\rm adv} + q_{\rm rad}^{\rm diff}. \label{eq-HC1}
\end{eqnarray}
Each term in this equation is defined as,
\begin{eqnarray}
  q_{\rm gas}^{\rm adv}&\equiv&-\left[\nabla\cdot(e\bm{v}) + p\nabla\cdot\bm{v}\right],\label{eq-HC2}\\
  q_{\rm rad}^{\rm adv}&\equiv&\nabla\cdot(E_{0}\bm{v}) + \nabla\bm{v}:\bm{{\rm P}}_{0},\label{eq-HC3}\\
  q_{\rm rad}^{\rm diff}&\equiv&\nabla\cdot\bm{F}_{0}.\label{eq-HC4}
\end{eqnarray}
Here,
$\Phi_{\rm vis}$ is the viscous heating,
$q_{\rm gas}^{\rm adv}$ is the advection of the gas including work by the gas pressure,
$\Gamma_{\rm comp}$ is the Compton heating and cooling,
$(4\pi\kappa B -c\kappa E_{0})$ is the net heating rate due to emission and absorption of radiation,
$q_{\rm rad}^{\rm adv}$ is the advection of the radiation including work by the radiation pressure,
and $q_{\rm rad}^{\rm diff}$ is the radiation diffusion.

Figure \ref{fig17} shows the angle-dependences of these heating and cooling rates at each radius.
We numerically confirmed equations (\ref{eq-HC0}) and (\ref{eq-HC1}) roughly hold.
See also table \ref{table2} for the summary of the dominant terms.

It is important to note that it is not advective cooling but advective heating that works in the gas energy equation.
The reason for this can be understood in the following way (Nakamura et al. 1997).
Entropy is generated via viscous dissipation within the gas so that the gas is heated.
But the internal energy is quickly transported to the radiation via the Compton process.
This results in a monotonic decrease of the gas entropy towards the center.
Since the higher-entropy gas moves inwards, the advection works as heating.

\begin{table*}[h]
  \tbl{Dominant heating and cooling rate}{
    \begin{tabular}{llllll}
      \hline    
      region 	&  angle	&     	& radius   & gas energy equation  & radiation energy equation\\
                & $(\theta)$    &       & $(r)$    & heating / cooling    & heating / cooling \\    
      \hline    
      equatorial plane & ($\sim90^{\circ}$) & inner part & ($r\lesssim 330~r_{\rm S}$)
      & $\Phi_{\rm vis}\sim \Gamma_{\rm comp}$               & $\Gamma_{\rm comp}\sim q_{\rm rad}^{\rm adv}$\\
      &	& middle part & ($330~r_{\rm S}\lesssim r \lesssim r_{\rm qss}$) 
      & $\Phi_{\rm vis}\sim (4\pi\kappa B -c\kappa E_{0})$  & $(4\pi\kappa B -c\kappa E_{0})\sim q_{\rm rad}^{\rm diff}$\\
      \hline
      disk surface & ($\sim\theta_{\rm surf}$)  &  & ($r\lesssim r_{\rm qss}$)         & $\Phi_{\rm vis}+q_{\rm gas}^{\rm adv}\sim \Gamma_{\rm comp}$ & $\Gamma_{\rm comp}\sim q_{\rm rad}^{\rm adv}+q_{\rm rad}^{\rm diff}$\\
      \hline    
  \end{tabular}}
  \begin{tabnote}
    The dominant heating and cooling rate in gas and radiation energy equations at each spatial point (see equation \ref{eq-HC0} and \ref{eq-HC1}).
  \end{tabnote}
  \label{table2}
\end{table*}

The dominant terms in equations (\ref{eq-HC0}) and (\ref{eq-HC1}) vary at each spacial position and is represented in table \ref{table2}.
When we focus near the equatorial plane ($\theta\sim 90^{\circ}$) in figure \ref{fig17},
the viscous heating is much larger than the advection of the gas ($\Phi_{\rm vis} \gg q_{\rm gas}^{\rm adv}$).
The energy is transported from gas to radiation by the Compton effect near the black hole,
whereas it is transported by the net radiation heating and cooling far from the black hole.
The advection of the radiation is dominant near the black hole,
but the radiation diffusion is dominant far from the black hole.
Hence, the energy balance near the equatorial plane is roughly $\Phi_{\rm vis} \sim \Gamma_{\rm comp} \sim q_{\rm rad}^{\rm adv}$ near the black hole,
and $\Phi_{\rm vis} \sim (4\pi\kappa B -c\kappa E_{0}) \sim q_{\rm rad}^{\rm diff}$ far from the black hole.

Let us next see the region just below the disk surface.
We then see that in figure \ref{fig17},
the viscous heating is comparable to advection of the gas ($\Phi_{\rm vis} \sim q_{\rm gas}^{\rm adv}$).
The energy is transported from gas to radiation by the Compton effect.
The advection of the radiation is comparable to the radiation diffusion.
The relation of equation (\ref{eq-HC0}) and (\ref{eq-HC1}) near the photosphere is roughly $\Phi_{\rm vis} +q_{\rm gas}^{\rm adv}\sim \Gamma_{\rm comp}  \sim q_{\rm rad}^{\rm adv}+q_{\rm rad}^{\rm diff}$.

The energy is carried by the advection of the radiation in vertical direction near the black hole ($R\lesssim R_{\rm trap}\sim \rtrapsph~r_{\rm S}$).
This process corresponds to energy transportation by the convection.
We confirmed that the advection term, $\nabla\cdot(E_{0}\bm{v})$ is larger than the work by radiation pressure, $\nabla\bm{v}:\bm{{\rm P}}_{0}$, by the factor of several(i.e. $q_{\rm rad}^{\rm adv}\sim \nabla\cdot(E_{0}\bm{v})$).
The advection term, $\nabla\cdot(E_{0}\bm{v})$, has radial and angular components,
and these components are approximately comparable to each other.
The radiation energy is carried by $E_{0}\bm{v}$ in vertical direction.
In other words, the radiation energy moves with gas.


We understand the radiation is trapped in gas by Lorentz transformation of the radiation flux (Ohsuga et al. 2007).
The formula in the optically thick region is written as,
\begin{eqnarray}
  F_{\rm lab}^{i}&=&F_{0}^{i}+v^{i}E_{0}+v_{j}{\rm P}_{0}^{ij}\propto v^{i}E_{0}. \label{eq-lorentz}
\end{eqnarray}
Here, $F_{\rm lab}^{i}$ is the radiation flux in the laboratory frame.
The comoving flux in the optically thick region is represented by the diffusion approximation, $\bm{F}_{0} = -c\nabla E_{0}/(3\rho\kappa)$,
and we assume that the radiation diffusion is inefficient, $F_{0}^{i}\ll v^{i}E_{0}$ (i.e. photon-trapping effect).
The third term on the right-hand-side of equation (\ref{eq-lorentz}),
$v_{j}{\rm P}_{0}^{ij}$, is equal to $E_{0}v^{i}/3$,
and corresponds to the work by the radiation pressure (Mihalas \& Mihalas, 1984).
From these relations,
we understand that $F_{\rm lab}^{i}\sim v^{i}E_{0}$ is established in the super-Eddington accretion disk,
and this formula means that the radiation moves with gas.
Especially,
energy is transported inward with gas velocity,
but the energy carrier is the radiation under the radiation pressure dominant region.

\section{Discussion}
\subsection{Comparison with the super-Eddington accretion model in Kitaki et al. (2018)}
\label{sec-ktk2018}
\begin{figure}[h]
\begin{center}
  \includegraphics[width=80mm,bb=0 0 360 288]{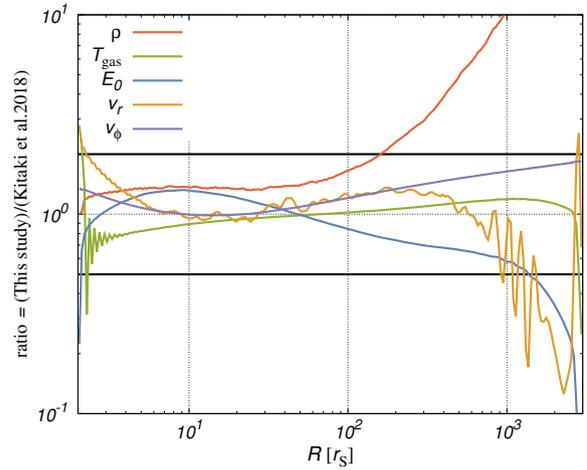}
\end{center}
\caption{
  The ratios of the values of some physical quantities (such as density, temperature, etc)
  obtained by the present simulation to those derived by the formulas of Kitaki et al. (2018).
  (The following values are inserted to the formulas:
  $M_{\rm BH}=10M_{\odot}$, $\dot{M}_{\rm BH}=\mdotinave~L_{\rm Edd}/c^{2}$, and $\alpha=0.1$.
  The colored lines represent the gas mass density (red), the gas temperature (green), the radiation energy density (blue), the radial velocity (orange), and the azimuthal velocity (purple).
  Both results agree well with each other (except for the density at large radii)
  within a factor of $2$ in the region sandwiched by the two horizontal black lines.
}
\label{fig10}
\end{figure}

Kitaki et al. (2018) performed rather extensive parameter studies of the RHD simulations of 
the super-Eddington accretion flow and derived semi-analytical formulas 
for representative physical quantities of the disk (e.g. temperature, density, radial velocity, and so on)
as functions of black hole mass, accretion rate, and radius.
Figure \ref{fig10} shows the ratios of the present numerical values to those reported by Kitaki et al. (2018).
We focus on the range $r\lesssim \rqss~r_{\rm S}$ where the quasi-steady state is achieved
(see section \ref{sec-mass-in-out}).
The present numerical results except for gas mass density agree well with those of Kitaki et al. (2018) within a factor of $2$.

As for the gas mass density we find a reasonable agreement in the inner region ($r\lesssim 200~r_{\rm S}$) but we notice significant discrepancies at larger radii.
The reason for the discrepancies can be understood in relation to the fact that 
the mass outflow rate is comparable to the (net) inflow rate there.
In fact, the inflow and outflow rates increase
outward at $r\gtrsim200~r_{\rm S}$ (see figure \ref{fig3}).
The formulas by Kitaki et al. (2018) were derived in the regime, in which outflow is negligible, compared with inflow.
Although the density profile shows some discrepancies, the radial velocity profile does not,
since the latter is rather insensitive to $\dot{M}_{\rm BH}$ (Watarai 2006; Kitaki et al. 2018).
In conclusion, we are not allowed to apply the formulas to the present numerical results in the region $r\gtrsim200~r_{\rm S}$ where the outflow is substantial.

As future work it will be useful to derive new formulas describing the accretion disk 
structure in the regime, in which the outflow rate is comparable to the inflow rate.

\subsection{Impact on the environments}
\begin{figure}[h]
\begin{center}
  \includegraphics[width=80mm,bb=0 0 360 648]{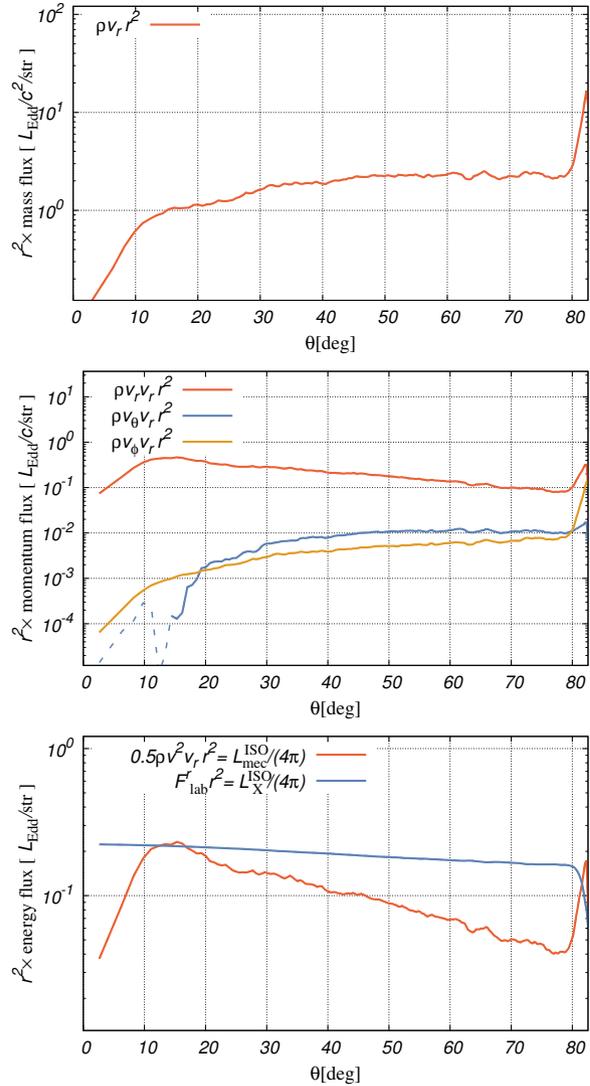}
\end{center}
\caption{
  The angular profile of the mass fluxes (top), the momentum fluxes (middle), and the energy fluxes (bottom) at $r=2545~r_{\rm S}$ in the range of $\theta=0^{\circ}-83^{\circ} (=\theta_{\rm surf})$.
  The solid lines mean that each value is positive (e.g. $0 < \rho v_{\theta}v_{r} \propto v_{\theta}$),
  and the dashed lines mean that each value is negative (e.g. $0 > \rho v_{\theta}v_{r} \propto v_{\theta}$).
}
\label{fig16}
\end{figure}

As we stressed in section \ref{sec-introduction}, the impact on the environments seems to have been
grossly overestimated in the previous simulation studies, since
the outflow was mostly launched from the initial Keplerian torus. 
In our simulation, we can give more realistic estimations on the impact by the outflow.

Figure \ref{fig16} shows the $\theta$ profiles of the mass flux, the momentum fluxes, and the energy fluxes measured at $r=2545~r_{\rm S}$ multiplied by $r^2$ in the upper to lower panels, respectively.
(In other words,
$d\dot{M}/d\Omega$, $d\dot{J}_{i}/d\Omega$ with $i = r$, $\theta$, and $\phi$,
and $d\dot{E}/d\Omega$ are plotted, 
where $\dot{J}_{i}$ is the momentum flow rate,
and $\dot{E}$ is the energy flow rate.)
The energy fluxes are also connected to the isotropic X-ray luminosity and the isotropic mechanical luminosity calculated by
\begin{eqnarray}
  L_{\rm X}^{\rm ISO}(\theta)&=& 4\pi r^{2} F_{\rm lab}^{r},\\
  L_{\rm mec}^{\rm ISO}(\theta)&=& 2\pi r^{2} \rho v^{2}v_{r}.
\end{eqnarray}
Here, $v^{2}=v_{r}^{2}+v_{\theta}^{2}+v_{\phi}^{2}$ is the total gas velocity,
and 
we assume that
radiation is emitted predominantly in the X-ray band,
since the ratio between the X-ray luminosity to the bolometric luminosity is $\sim71$--$98\%$ from Kitaki et al. (2017) and Narayan et al. (2017).

We notice that the radiation luminosity has a peak in the face-on ($\theta=0$) direction,
while the mass outflow rate increases towards the edge-on direction, and
the momentum flux and the mechanical luminosity reach their maximum at $\theta \sim 15^\circ$.
In the inner funnel region (with $\theta\sim0^{\circ}-15^{\circ}$),
all the lines except for the radiative flux rapidly decrease with a decrease of $\theta$.
This is because the gas mass density rapidly decreases inward (towards the rotation axis).
Although the radial velocity increases towards the rotation axis, 
the change in $\rho$ overcomes that in $v_r$ so that $\rho v_r$ decreases with decreasing $\theta$.

In the outflow region ($\theta\sim20^{\circ}-83^{\circ}$), by contrast,
the radial components of the fluxes ($\rho v_{r}$, $\rho v_{r}^{2}$, $0.5\rho v^{2}v_{r}$) only slightly or scarcely change in the polar direction.

The isotropic X-ray luminosity decreases slightly with increasing angle by less than factor 2.
The previous studies showed, however, that the isotropic X-ray luminosity varies by a factor of $\sim10$ in the azimuthal direction (e.g., Ogawa et al. 2017).
This difference stems from the different outflow properties.
That is, much larger amount of gas is blown away in the previous studies
because of the small value of the Keplerian radius being adopted.
It is difficult for
photons generated inside the disk region (with small $\theta$) to pass through the outflow region (at large $\theta$).

Observational determination of the inclination angle seems to be difficult.
By contrast, the spectral shape is sensitive to the inclination angle because of the significant Compton scattering within the outflow region (Kawashima et al. 2012).
Spectral calculation based on our hydrodynamic simulation data is left as future work.

\subsection{The energy conversion}
The energy conversion efficiency is one of the key quantities in accretion problems.
In subsection \ref{sec-mass-in-out},
we evaluated that
the inflow rate at the black hole is $\dot{M}_{\rm BH}\sim\mdotinave ~L_{\rm Edd}/c^{2}$,
and that the outflow rate above the surface of the disk at around $r_{\rm out}$ is $\dot{M}_{\rm outflow}\sim\mdotoutend~L_{\rm Edd}/c^{2}$.
We should note, that the numerical results in $r\geq r_{\rm qss}=\rqss ~r_{\rm S}$ are not so reliable,
since the quasi-steady assumption does not hold there.
But we can accurately estimate the outflow quantities, such as the outflow rate $\dot{M}_{\rm outflow}$ at $r_{\rm out}$, since the outflows are launched at relatively small radii; i.e., $r_{\rm lau}\sim\rlau ~r_{\rm S}< r_{\rm qss}$.
The inflow and outflow conversion efficiencies are calculated as,
\begin{eqnarray}
  \beta&\equiv&\frac{\dot{M}_{\rm outflow}}{\dot{M}_{\rm BH}}\sim0.14,\\
  \beta_{\rm in}&\equiv&\frac{\dot{M}_{\rm BH}}{\dot{M}_{\rm BH}+\dot{M}_{\rm outflow}}\sim0.88,\label{eq-ene1}\\
  \beta_{\rm out}&\equiv&\frac{\dot{M}_{\rm outflow}}{\dot{M}_{\rm BH}+\dot{M}_{\rm outflow}}\sim0.12. \label{eq-ene2}
\end{eqnarray}
Here, the denominators of equations (\ref{eq-ene1}) and (\ref{eq-ene2}),
$\dot{M}_{\rm BH}+\dot{M}_{\rm outflow}$,
mean the injected mass flow rate from surrounding environment under the assumption that the net flow rate is entirely constant in radius.
Hence, we conclude that about 12\% of the injected gas turns into outflow.

The luminosity measured by a distant observer is calculated by
\begin{eqnarray}
  L_{\rm X}&\equiv&4\pi\int_{0}^{\theta_{\rm surf}} d\theta\sin\theta~r^{2}~{\rm max}\left\{F_{\rm lab}^{r},0\right\},
\end{eqnarray}
and is $L_{\rm X}\sim 2.5~L_{\rm Edd}$ at $r = 2545 ~r_{\rm S}$ (near the outer boundary).
While, the predicted luminosity from slim disk formula is given by (Watarai et al. 2001),
\begin{eqnarray}
  L_{\rm slim}&=&\left[1+\ln\left(\frac{1}{30}\frac{\dot{M}_{\rm BH}}{L_{\rm Edd}/c^{2}}\right)\right]L_{\rm Edd}\sim2.8~L_{\rm Edd},
\end{eqnarray}
and is in a reasonable agreement.

The mechanical luminosity of the outflow is given by
\begin{eqnarray}
  L_{\rm mec}&\equiv&4\pi\int_{0}^{\theta_{\rm surf}} d\theta\sin\theta~r^{2}~{\rm max}\left\{\frac{1}{2}\rho v^{2}v_{r},0\right\},
\end{eqnarray}
and is $L_{\rm mec}\sim0.16~L_{\rm Edd}$ at $r = 2545 ~r_{\rm S}$ (near the outer boundary).
The ratio of the luminosities is $L_{\rm mec}/L_{\rm X}\sim0.07$.
Hence, the energy carried outside by the radiation is larger than that by the outflow.

\subsection{Outflow in ULXs}
\begin{table*}[h]
  \tbl{The X-ray Luminosities and the mechanical luminosities}{
  \begin{tabular}{llll}
    \hline
    object & $L_{\rm X}^{\rm ISO}[10^{39}{\rm erg/s}]$ & $L_{\rm mec}[10^{39}{\rm erg/s}]$ & $L_{\rm mec}/L_{\rm X}^{\rm ISO}$\\
    \hline
    our simulation ($\theta = 2.6^{\circ}$--$80^{\circ}$) & $\sim2.6$--$3.7$ & $\sim0.20$ & $\sim0.05$--$0.08$\\
    \hline
    Holmberg II X-1 (ULX)& $\sim5$--$16$ \footnotemark[$*$]    & $\sim0.7$ \footnotemark[$\dagger$] & $\sim0.04$--$0.14$\\
    IC342 X-1 (ULX)     & $\sim10$--$20$ \footnotemark[$\ddagger$]  & $\sim3$ \footnotemark[$\dagger$] & $\sim0.15$--$0.3$ \\
    \hline
  \end{tabular}}
  \begin{tabnote}
    Here,
    $L_{\rm X}^{\rm ISO}$ is the isotropic X-ray luminosity of a central object,
    and
    $L_{\rm mec}$ is the mechanical luminosity of the outflow.
    \footnotemark[$*$] Kaaret et al. (2004), 
    \footnotemark[$\dagger$] Abolmasov et al. (2006),
    \footnotemark[$\ddagger$] Shidatsu et al. (2017).
  \end{tabnote}
  \label{table3}
\end{table*}

Some ULXs are accompanied with the optical nebula with extents of $10$--$100$ parsecs (e.g. Kaaret et al. 2004) and/or radio bubble with extents of $10$--$100$ parsecs (e.g. Berghea et al. 2020).
These nebulas are thought to originate from the outflow in super-Eddington accretion flow.
The isotropic X-ray luminosities of the central object and the mechanical luminosities of the outflow are listed in table \ref{table3} for some ULXs.
On the observational side,
the isotropic X-ray luminosities are evaluated from the photons which come to observer directly,
and the mechanical luminosities are evaluated from the optical radiation.
The values of $L_{\rm X}^{\rm ISO}$ and $L_{\rm mec}$ evaluated in the present study are thus consistent with the observed luminosities in ULXs (see table \ref{table3}).

We wish also to note that the iron absorption lines in the X-ray with Doppler velocities of $\sim 0.2~c$ are discovered by the stacking analysis (Pinto et al. 2016),
and these velocities are consistent with our results.


\subsection{Future issues}
The M-1 closure method, alternative method to calculate radiation flux etc, is extensively used in recent simulations instead of the FLD method (e.g., S\c{a}dowski et al. 2015).
We, however, note that there will be practically no big differences in the calculated accretion disk structure between them,
since both methods give the correct formula of the radiative diffusion approximation in the optically thick regime.
In the optically thin region (e.g., outflow region), by contrast,
we find slight differences,
since the M-1 closure give the different solution compared with the FLD method there (e.g., in the beam problem, the shadow test, etc; see Gonz\'{a}lez et al. 2007).
Further, we wish to point, however, that the M-1 closure method does not always give the correct radiation fields,
since the M-1 closure method is also an approximation
and it is known to produce inaccurate results in the nonuniform radiation fields
(see, e.g., Ohsuga \& Takahashi 2016 for the case of a radiation hydrodynamic shock).
This issue will only be resolved by full-transfer simulations in future.

We assume the equatorial symmetry in our simulation.
When we expand the simulation box in the polar direction from $\theta=0$--$\pi/2$ to $\theta=0$--$\pi$,
the flow pattern in the disk may be slightly modified by the convective motion across the equatorial plane,
which is seen in S\c{a}dowski et al. (2016).
But, it is hard to believe that the disk structure will change dramatically from our results.

The next issue is three dimensional simulations.
S\c{a}dowski et al. (2016) compared the results by 2D and 3D GR-RMHD simulations of the super-Eddington accretion flow,
and reported that the physical values in the accretion disk (e.g., rotational velocity, surface density) are almost same within the error of $10\%$.
They also showed that the radiation energy density differs only within the factor of 2 between 2D and 3D simulations.
We thus believe that there will be no significant differences,
as long as the global structure is concerned,
but 3D simulations are definitely needed in future for more advanced study,
e.g., spatio-temporal variation studies.

We used the $\alpha$ prescription for simplicity (Shakura \& Sunyaev 1973),
although the magnetorotational instability (MRI) is believed to be one of the most plausible origins of the viscosity (Balbus \& Hawley 1991, 1992).
S\c{a}dowski et al. (2015) calculated the super-Eddington accretion flow by GR-RMHD simulation
and found that the viscosity parameter is $\alpha \sim 0.1$ and is roughly constant in the radius outside the innermost stable circular orbit (ISCO).
Therefore,
we conclude that our assumption ($\alpha = 0.1$) is reasonable,
but in future we need full radiation-MHD simulations for confirmation.

\section{Concluding remarks}
\begin{figure}[h]
\begin{center}
  \includegraphics[width=80mm,bb=0 0 1024 768]{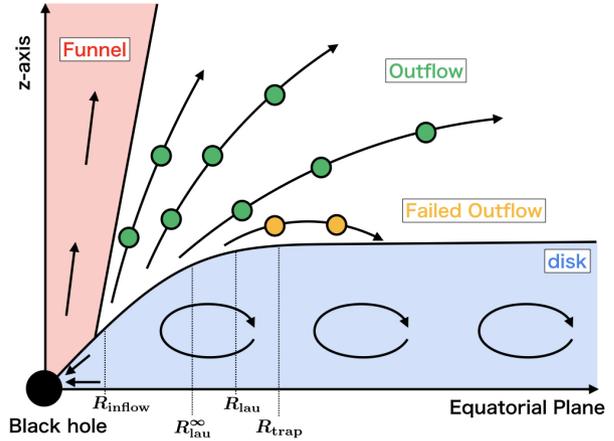}
\end{center}
\caption{
  Schematic view of the structure of the super-Eddington accretion flow and associated outflow based on our numerical results.
  The black arrows indicate the gas motion.
}
\label{figpon}
\end{figure}

In the present study we have carefully solved the structure of the super-Eddington accretion flow
and associated outflow and examined where in the accretion flow the outflow is launched 
and how much material, momentum, and energy can be blown away to a large distance.
For this purpose, we grossly expand the simulation box to $r_{\rm out} = 3000~r_{\rm S}$ 
and set an extremely large initial Keplarian radius, $r_{\rm K}= 2430~r_{\rm S}$,
compared with those adopted in the previous studies.
We could, hence, achieve the unprecedentedly large quasi-steady radius, $r\leq r_{\rm qss}\sim 600~r_{\rm S}$, within which a quasi-steady state is realized.

In figure \ref{figpon} we summarize our view of the structure of the super-Eddington accretion flow and associated outflow obtained through the present study.
In this figure we plot gas motion (both of circular motion within the disk and outflow from the disk surface).
The main features can be summarized as follows:

\begin{itemize}
\item
The disk thickness ($H$) is roughly proportional to $R$ (i.e., $H/R \sim 1$) near the black hole,
whereas it is constant in radius, $H \sim (2$--$3)\times 10^{2}~r_{\rm S}\sim (1.1$--$1.7)\dot{m}_{\rm BH}r_{\rm S}$, far from the black hole.
The photon-trapping radius, $R_{\rm trap}\sim\rtrapf~r_{\rm S}\sim2.5~\dot{m}_{\rm BH}r_{\rm S}$, approximately separates these two regions.
This feature is consistent with the prediction of the slim disk model and the standard Shakura-Sunyaev disk.

\item
From the streamline analysis we find that 
the gas on the disk surface at $R\leq R_{\rm lau}\sim\rlaucyl~r_{\rm S}\sim1.3~\dot{m}_{\rm BH}r_{\rm S}$ is blown away
to produce outflow (see figure \ref{fig6}).
The genuine outflow (which goes to reach the outer calculation boundary) is launched at $R\lesssim R_{\rm lau}^{\infty}\sim \rlauinfcyl ~r_{\rm S}\sim0.75~\dot{m}_{\rm BH}r_{\rm S}$, well inside the trapping radius,
while the failed outflow originates from the region between $R\sim R_{\rm lau}^{\infty}$ and $R_{\rm lau}$ (see figure \ref{fig6}).
The systematic simulation studies to confirm the $\dot{m}_{\rm BH}$-dependences of these radii (e.g., $R_{\rm lau}$) are left as future work.

\item
  The black hole accretion rate in our study is $\dot{M}_{\rm BH}\sim \mdotinave ~L_{\rm Edd}/c^2$
  and the outflow rate is $\dot{M}_{\rm outflow}\sim \mdotoutend~L_{\rm Edd}/c^2$.
  The ratio of the isotropic X-ray luminosity to the mechanical luminosity is $L_{\rm mec}/L_{\rm X}^{\rm ISO} \sim 0.05$--$0.08$, which is consistent with the observations of ULXs surrounded by optical nebulae.

\item 	
We separately examined the energy balance for the radiation and for the gas.
Around the equatorial plane,
the gas is heated via the viscous dissipation and is cooled by emitting radiation at large radii, as was formulated in the standard disk model,
but is cooled by the inverse Compton scattering at small radii.
The radiation is heated through the inverse Compton scattering
and is cooled by the advection of the radiation at small radii,
as was formulated in the slim disk model.
The situation is somewhat different near the disk surface,
where gas is heated both by the viscous dissipation and the advective heating (not cooling),
which occurs since the entropy decreases inward as gas accretes as a consequence of significant Compton cooling.

\item
Convection (or large-scale circulation) occurs nearly entirely in the accretion disk.
The convective motions are observed even inside $R_{\rm inflow}\sim 40~r_{\rm S}\sim0.22~\dot{m}_{\rm BH}r_{\rm S}$ in snapshots (see figure \ref{fig113}),
although they are smeared out when time-averaged (see figure \ref{fig1_small}).
The direction of the convective motion is sometimes clockwise and sometimes anti-clockwise,
but more frequently we see clockwise motion,
when we set the black hole on the lower-left corner (see figures \ref{fig13} and \ref{fig113}).

\item
Most of the previous simulation studies show a puffed up structure near the black hole ($R\ll R_{\rm trap}$),
which seems to be formed as a direct consequence of adopting a very small Keplerian radius.
We should make caution that large amount of mass outflow can be produced by such a puffed up structure
and that the cases with small Keplarian radii may be applied to the tidal disruption events which undergoes super-Eddington accretion.
This is because the disrupted objects by tidal action came close to the black hole.
\end{itemize}

\begin{ack}
  Numerical computations were mainly carried out on Cray XC50 and the analysis servers at Center for Computational Astrophysics, National Astronomical Observatory of Japan.
  This work is supported in part by JSPS KAKENHI Grant Numbers,
  17H01102 (K.O.), 18K03710 (K.O.), 18K13594 (T.K.), 19J14724 (T.K.), 20K04026 (S.M.),
  and
  is also supported by MEXT as “Program for Promoting Researches on the Supercomputer Fugaku”
  (Toward a unified view of the universe: from large scale structures to planets, K.O., T.K.) and by Joint Institute for Computational Fundamental Science (JICFuS, K.O.).
\end{ack}


\end{document}